\documentclass[aps,showpacs,preprintnumbers,amsmath,amssymb]{revtex4}
\usepackage{graphicx}
\usepackage{bm}

\newcommand{\lb}{\left}
\newcommand{\rb}{\right}
\newcommand{\Complex}{\mathbb{C}}
\newcommand{\Real}{\mathbb{R}}
\newcommand{\Natural}{\mathbb{N}}

\newcommand{\Tr}{\mathop{\mathrm{Tr}}}
\newcommand{\extr}{\mathop{\mathrm{Extr}}}
\newcommand{\argmax}{\mathop{\mathrm{argmax}}}
\newcommand{\argmin}{\mathop{\mathrm{argmin}}}
\newcommand{\calbmP}{{\cal {P \hspace{-2.9mm} P}}}
\renewcommand{\Re}{\mathrm{Re}}
\renewcommand{\Im}{\mathrm{Im}}
\newcommand{\Rr}{\bm{R}_{\mathrm{r}}}
\newcommand{\Rt}{\bm{R}_{\mathrm{t}}}
\newcommand{\arctanh}{\mathrm{arctanh}}

\begin{document}

\title {Statistical mechanical analysis of the Kronecker channel model
 for MIMO wireless communication}

\author{Atsushi Hatabu}
\affiliation{
System IP Core Research Laboratories, NEC Corporation,
Kawasaki 216-8555, Japan and \\
Department of Computer Intelligence and Systems Science,
Tokyo Institute of Technology,
Yokohama 226-8502, Japan
}

\author{Koujin Takeda}
\author{Yoshiyuki Kabashima}
\affiliation{
Department of Computer Intelligence and Systems Science,
Tokyo Institute of Technology,
Yokohama 226-8502, Japan
}
\date{September 18, 2009}

%%%%%%%%%%%%%%%%%%%%%%%%%%%%%%%%%%%%%%%%%%%%%%%%%%%%%%%%%%%%%%%%%%%
%%%%%%%%%%%%%%%%%%%%%%%%%%%%%%%%%%%%%%%%%%%%%%%%%%%%%%%%%%%%%%%%%%%
\begin{abstract}

The Kronecker channel model of wireless communication is analyzed using 
statistical mechanics methods. In the model, spatial proximities among 
transmission/reception antennas are taken into account as certain correlation
matrices, which generally yield non-trivial dependence among symbols
to be estimated. This prevents accurate assessment of the
communication performance
by na\"{i}vely using a previously developed analytical scheme based on a matrix integration 
formula. 
In order to resolve this difficulty, we develop a formalism that can formally 
handle the correlations in Kronecker models based on the known 
scheme. Unfortunately, direct application of the developed scheme is, 
in general, practically difficult. However, the formalism is still useful, indicating that 
the effect of the correlations generally increase after the fourth order 
with respect to correlation strength. Therefore, the known
analytical scheme offers a good approximation in 
performance evaluation when the correlation strength is
 sufficiently small. For a class of specific correlation, we show that
the performance analysis can be mapped to the problem of one-dimensional
spin systems in random fields, which can be investigated
without approximation by the belief propagation algorithm.

\end{abstract}

\pacs{84.40.Ua, 75.10.Nr, 89.70.-a}

\maketitle

%%%%%%%%%%%%%%%%%%%%%%%%%%%%%%%%%%%%%%%%%%%%%%%%%%%%%%%%%%%%%%%%%%%
%%%%%%%%%%%%%%%%%%%%%%%%%%%%%%%%%%%%%%%%%%%%%%%%%%%%%%%%%%%%%%%%%%%

\section{Introduction}
\label{sec:intro}

Recently, in the field of information science, techniques for
efficiently handling systems with large amounts of data 
are strongly required, and statistical
mechanics have attracted a great deal of attention.
The number of applications in information science
to which analytical schemes in statistical mechanics can be applied 
is increasing, and such applications offer 
a variety of consequences \cite{bib:Nishimori}, 
some of which are not possible by standard techniques in information science.
Information processing is a notable example.

In the present study, we investigate wireless communication systems.
Multiple\,-\,input multiple\,-\,output (MIMO) systems and 
code division multiple access (CDMA) systems in wireless communication
have mathematical structures that are similar to those of disordered spin systems in
physics, and analytical tools in statistical mechanics,
 such as the replica method and 
mean field approximations, have enabled 
performance analysis and improved processing algorithms
for actual communication systems
\cite{bib:TanakaCD1,bib:TanakaCD2,
bib:KabashimaDecoding,bib:Guo2005,bib:CDMAo1,bib:NeirottiSaad,bib:EYSK2009}.
In these studies, the communication process is described by a linear equation 
with transmitted signals $\bm{b}\in\Complex^{K}$ and received signals
$\bm{r}\in\Complex^{L}$ 
using an $L \times K$ channel matrix
$\bm{H}=(H_{lk})\in\Complex^{LK}$ and noise
$\bm{\eta}\in\Complex^{L}$ as 
\begin{eqnarray}
\label{eq:model}
\bm{r}&=&\bm{H}\bm{b}+\sigma\bm{\eta},
\end{eqnarray}
where $\sigma^2$ describes the noise power. 
Throughout the present paper, matrices and vectors are denoted in bold. 
In the above equation, the dimension $K$ represents the number of
multiple transmission antennas in the MIMO system, 
whereas $L$ corresponds to the number of reception antennas. 
Clarifying the feature of the above-mentioned
communication channel by the standard method of information theory is 
technically difficult because of the randomness in $\bm{H}$ 
and the discreteness of variable $\bm{b}$.
However, statistical mechanical analysis enables us to avoid such difficulties 
in the limit of infinite system size. 

In previous studies based on statistical mechanical methods,
channel matrix $\bm H$ was characterized by a property that the 
cross correlation $\bm H^\dagger \bm H$ can be handled as a typical sample from a
rotationally invariant matrix ensemble, as follows:
\begin{equation}
\label{eq:basic_class}
 \bm H^\dag \bm H = \bm U\bm D\bm U^\dag,
\end{equation} 
where $\bm U$ is a sample randomly chosen from the uniform distribution of
$K$-dimensional unitary matrices,
 and $\bm D$ is a $K$-dimensional diagonal matrix.
If $\bm D$ has an asymptotic and deterministic eigenvalue
distribution $\rho_{\bm D}(\lambda)$ with a large matrix size limit
 $L,K \rightarrow \infty$ while keeping $\beta \equiv K/L$ finite,
the features of this channel can be characterized by
$\rho(\lambda)$, in conjunction with the replica method, and 
the performance of the channel can be assessed
\cite{bib:TUK2006,bib:THK2007,bib:mimo1}
by a matrix integration formula
\cite{bib:gfunc1, bib:gfunc2, bib:gfunc3, bib:freeprob,
 bib:TulinoVerdu}, which is defined for $\rho(\lambda)$.

However, one problem remains. For the simplest case in which each element
of random matrix $\bm H$ is drawn from an 
independent and identically-distributed (i.i.d.)
Gaussian distribution, the property of rotational invariance concerning 
the cross correlation is satisfied.
However, such property does not necessarily hold for general matrix ensembles of MIMO systems.
For instance, in the Kronecker model \cite{bib:kronecker1},
which is one of the standard models in the theory of wireless communication,
the elements of the channel matrix are not drawn from an 
i.i.d Gaussian distribution, but are instead drawn from an $L \times K$-dimensional joint Gaussian distribution. More precisely, the channel matrix $\bm H$ is described as 
\begin{equation}
\label{eq:model_K}
\bm{H}=\sqrt{\Rr}\bm{\Xi}\sqrt{\Rt},
\end{equation}
where each component of an $L \times K$ rectangular matrix
$\bm\Xi=(\Xi_{lk})$
is drawn from a complex i.i.d. Gaussian distribution:
$P(\Xi_{lk})=L\pi^{-1}e^{-L|\Xi_{lk}|^2}$($1\leq l\leq L, 1\leq k\leq K$).
$\Rr\in\Complex^{L^2}$ and $\Rt\in\Complex^{K^2}$ are $L$- and $K$-dimensional
deterministic matrices, which are Hermitian and indicate correlations among reception antennas and transmission antennas, respectively. 
In a previous study\cite{bib:THK2007}, we analyzed this system by means of
the matrix integration formula. However,
for this system, the matrix ensemble is not rotationally invariant and, 
accordingly, the result of performance analysis via the matrix integration 
formula may not hold exactly.

One of the goals of the present paper is to
develop a scheme that can handle the dependence on $\sqrt{\Rr}$ and
 $\sqrt{\Rt}$ in Eq. (\ref{eq:model_K}) explicitly.
In other words, the method developed herein relies on the direct 
integration of each matrix element in Gaussian random matrix $\bm \Xi$.
The results of analysis for mutual information indicate that
the roles of the deterministic matrices
 $\sqrt{\Rr}$ and $\sqrt{\Rt}$ are different. As will be shown later herein, the dependence on $\sqrt{\Rr}$ can be treated using the matrix integration technique, whereas the dependence on $\sqrt{\Rt}$ must be handled more carefully. The developed scheme can also be used to construct a practical
demodulation algorithm.
Another goal is to compare the performance of 
the Kronecker channel (\ref{eq:model_K}) via a novel analysis
with the performance of the matrix integration formula applied to
the entire cross-correlation matrix $\bm H^{\dagger} \bm H$,
as we demonstrated in Reference \cite{bib:THK2007}.
The two formulations are found to yield different 
result, which means that the application of 
the matrix integration formula to the cross-correlation matrix
$\bm H^{\dagger} \bm H$, in general, does not yield correct results.
However, when correlation among transmission
antennas, or when the off-diagonal element of the deterministic matrix
$\sqrt{\Rt}$ is sufficiently small, discrepancy between results of 
the scheme developed herein and that based on the matrix integration 
formula increases only after the fourth order with respect to 
correlation strength, implying that the formulation based on entire 
matrix integration yields good approximate results.
This suggests that although the matrix-integration technique is
generally an approximation, 
this technique is practically useful when the correlation is small,
because the matrix integration method enables 
the system to be characterized using only a few macroscopic variables, 
which significantly reduces the computational cost for analysis.

The remainder of the present paper is organized as follows. 
In Section \ref{sec:analysis}, we provide basic tools for performance analysis
and propose a novel approach to analyze the Kronecker channel model.
The analytical results differ from those obtained 
by matrix integration. In addition, we compare two results by a method of perturbative expansion
with respect to the correlation parameter, and a discrepancy appears
in the fourth-order coefficient of the correlation parameter,
which indicates that the discrepancy is small when the correlation is small.
In Section \ref{sec:algo}, we show that 
the demodulation algorithm can be constructed from the minimization
 scheme of Gibbs free energy without knowledge of the matrix integration.
We present the experimental results of the demodulation algorithm for
the Kronecker channel in Section \ref{sec:exp}.
As a special case, we consider a system with a tridiagonal form
of $\Rt$, where the analytical scheme
can be used for the random-field Ising chain. The results of a numerical experiment confirm the validity and usefulness of the proposed scheme. The final section presents a summary of the present paper. 

%%%%%%%%%%%%%%%%%%%%%%%%%%%%%%%%%%%%%%%%%%%%%%%%%%%%%%%%%%%%%%%%%%%
%%%%%%%%%%%%%%%%%%%%%%%%%%%%%%%%%%%%%%%%%%%%%%%%%%%%%%%%%%%%%%%%%%%

\section{Analysis} 
\label{sec:analysis}

Let us start with the communication channels described by Eq. (\ref{eq:model}).
For the noise, we assume that $\bm\eta$ is drawn from a white normal complex
Gaussian distribution $P(\bm\eta)=\pi^{-L}e^{-|\bm\eta|^2}$. 
Each component of the $K$-dimensional transmit vector $\bm b$
is generated from an i.i.d. information source and modulated.
For simplicity, the modulated components $b^0_k$ are quantized
to one of the elements in a set $\cal B$. 
For instance, for $S$-phase shift keying modulation
${\cal B} \equiv \{e^{2\pi is/S}\} (s=0,1,...,S-1)$. 
As special and well-known cases, 
${\cal B}\equiv \{\pm 1\}$ for binary-phase shift keying (BPSK) 
modulation, and ${\cal B} \equiv \{\pm 1/\sqrt{2}\pm i/\sqrt{2}\}$ for
quadrature-phase shift keying (QPSK) modulation.
The prior of the transmit vector is denoted by $P(\bm
b)=\prod_{k=1}^{K}P(b_k)$. Here $P(b_k)=1/|{\cal B}|$, and $|{\cal B}|$ is
the number of elements in ${\cal B}$. 

As mentioned in the introduction, we investigate the Kronecker model described by
the matrix of Eq. (\ref{eq:model_K}).
In order to apply statistical mechanical
schemes to the analysis of communication systems, we allow 
the number of antennas $L$ and $K$ to be sufficiently large
while keeping $\beta=K/L$ finite. 
Next, let us assume that for the matrices
$\Rr$ and $\Rt$ that there exist deterministic distributions
$\rho_{\Rr}(\lambda)$ and $\rho_{\Rt}(\lambda)$,
respectively, in the limit of infinite number of antennas. In addition, we 
assume that both distributions have compact supports and finite
moments, which affects the applicability of the matrix integration formula.

In the following, we consider only the case in which the receivers know the channel matrix $\bm{H}$ and the noise power $\sigma^2$ in advance. 
The performance of the communication channels can be analyzed
by estimating the mutual information between transmitted signals
$\bm b$ and the received signals $\bm r$, denoted by ${\cal I}_{\bm H}$. 
For MIMO systems, we have 
\begin{eqnarray}
\label{eq:mutualinfo}
{\cal I}_{\bm H} &=& -\frac{1}{K}\int_{\Complex^L}
 d\bm r  Z(\bm r)\ln Z(\bm r) - \frac{1}{\beta} \ln (\pi \sigma^2) 
- \frac{1}{\beta},
\nonumber \\
{\rm where}\ \ Z(\bm r) &\equiv & 
\Tr_{\bm b}P(\bm b)\frac{1}{(\pi\sigma^2)^L}
\exp\lb[-\frac{|\bm r-\bm H \bm b|^2}{\sigma^2}\rb].
\end{eqnarray}
In this article we use nat unit for mutual information and entropy.
In statistical mechanics, $Z(\bm r)$ serves as a partition
function, which depends on quenched randomness $\bm H$,
and ${\cal I}_{\bm H}$ is considered to represent the free energy.   

Following the standard technique, we use the replica
method to take the average over the channel matrix $\bm H$
in the mutual information:
\begin{eqnarray}
\label{eq:rep2}
\overline{{\cal I}_{\bm H}}
&=&
-\lim_{n\to 0}\frac{\partial}{\partial n} 
\frac{1}{K}\ln
 \overline{\int_{\Complex^L} d\bm r Z^{n+1}(\bm r)}
- \frac{1}{\beta} \ln (\pi \sigma^2) - \frac{1}{\beta},
\end{eqnarray}
where $\overline{\cdots}$ denotes averaging over
the distribution of channel matrix $\bm H$. 
For $n=0,1,2,\ldots$, we have  
\begin{eqnarray}
\label{eq:ffff1}
\int_{\Complex^L} d\bm r
Z^{n+1}(\bm r)
&=& 
\int_{\Complex^L} d\bm r 
\prod_{a=0}^{n} \left(  \Tr_{\bm b^a} P(\bm b^a) 
\frac{1}{(\pi\sigma^2)^L}
\exp{\lb[-\frac{|\bm r-\bm H \bm b^a|^2}{\sigma^2}\rb]}
\right)
\nonumber \\
&=& 
\left( \frac{\pi\sigma^2}{n+1} \right)^{L}
\left( \prod_{a=0}^{n} \Tr_{\bm b^a} P(\bm b^a) \frac{1}{(\pi\sigma^2)^{L}}\right)
\exp{\lb[-\frac{1}{\sigma^2}\Tr\lb(\bm H^\dag\bm H \bm L
\rb) \rb]},
\end{eqnarray}
where the $K$-dimensional square matrix $\bm L$ is defined by
$L_{kk'} \equiv
\sum_{a=0}^{n} b^{a}_{k} b^{a *}_{k'}
- \sum_{a b} {b}^{a}_{k} {b}^{b *}_{k'} / (n+1)$.
This can be rewritten as
$L_{kk'} = \sum_{a b} b^a_{k} {\cal P}^{ab} b_{k'}^{b *}$
by introducing the $(n+1)$-dimensional projection
matrix ${\cal P}^{ab} \equiv \delta^{ab} - 1 / (n+1)$.
Substituting channel matrix $\bm H$, given in Eq. (\ref{eq:model_K}), for 
the Kronecker model and integrating
with respect to $\bm\Xi$, we obtain 
\begin{eqnarray}
\label{eq:term21}
\overline{
\int_{\Complex^L} d\bm r
 Z^{n+1}(\bm r)}
&\propto&
\left( \prod_{a=0}^{n} \Tr_{\bm b^a} P(\bm b^a) \right)
\det \left( \bm I_{LK}+\frac{1}{\sigma^2 L}\Rr\otimes\sqrt{\Rt}\bm
L \sqrt{\Rt} \right)
\nonumber \\
&=&
\left( \prod_{a=0}^{n} \Tr_{\bm b^a} P(\bm b^a) \right)
\exp \left[ L \int d\lambda \rho_{\Rr} (\lambda)
 \Tr \ln \left( \bm I_K + \frac{\lambda}{\sigma^2 L}
 \sqrt{\Rt}\bm L \sqrt{\Rt} \right)
\right]
\nonumber \\
&=& \left( \prod_{a=0}^{n} \Tr_{\bm b^a} P(\bm b^a) \right)
\exp \left[ K  \Tr G_{\bm\Xi^\dag\Rr\bm\Xi}
\left( - \frac{1}{\sigma^2 K} \sqrt{\Rt}\bm L \sqrt{\Rt} \right)
\right]
\nonumber \\
&=& \int d\bm Q
\exp \lb[
K \Tr G_{\bm\Xi^\dag\Rr\bm\Xi}
\lb(-\frac{1}{\sigma^2} \calbmP \bm Q\rb) + \ln \Pi^{(n)} (\bm Q) \rb],
\end{eqnarray}
where $\bm I_{D}$ is the $D$-dimensional identity matrix,
$\bm Q$ is an $(n+1)$-dimensional matrix,
and $\otimes$ represents the Kronecker or direct product.
Note that the trace in Eq. (\ref{eq:term21}) is
$K$-dimensional in the second and third lines
and is $(n+1)$-dimensional in the last line. 
In the equation above, the following functions are defined:
\begin{eqnarray}
\Pi^{(n)}(\bm Q)&\equiv&
\lb\{\prod_{a=0}^{n} \Tr_{\bm b^a} P(\bm b^a)\rb\}
\lb\{\prod_{a=0}^{n} \delta(\bm b^{a\dag}\Rt\bm b^{a}-KQ_{aa})\rb\}
\lb\{\prod_{a<b}^{n}\delta(\bm b^{a\dag}\Rt\bm b^{b}-KQ_{ab})\rb\},
\\
\label{eq:Gfunc}
G_{\bm \Xi^{\dagger} \Rr \bm \Xi}(\bm A)
&\equiv& \frac{1}{\beta}\int d\lambda \rho_{\Rr}(\lambda)
\ln\lb(\bm I -\beta \lambda \bm A \rb),
\end{eqnarray} 
where $\rho_{\Rr}(\lambda)$ in the function $G_{\bm \Xi^{\dagger} \Rr \bm
 \Xi}$ is the eigenvalue distribution
for the matrix $\Rr$.
The function $G_{\bm \Xi^{\dagger} \Rr \bm \Xi}$ is the function obtained from
the matrix integration formula over unitary matrix Haar measure $d\bm U$
\cite{bib:gfunc1,bib:gfunc2,bib:gfunc3},
\begin{equation}
\label{eq:Gfuncdef}
\int d \bm U \exp \left( \Tr {\bm \Xi^{\dagger} \Rr \bm \Xi} 
{\bm A} \right)
\simeq \exp \left( {K \Tr G_{\bm \Xi^{\dagger} \Rr \bm \Xi}
\left( {\bm A}/K \right)} \right).
\end{equation}
Here, unitary matrix 
$\bm U$ to be integrated is defined
as a $K$-dimensional unitary matrix that diagonalizes
the random matrix product as $\bm \Xi^{\dagger} \Rr
\bm \Xi = \bm U^{\dagger} \bm D \bm U$,
where $\bm D$ is a diagonal matrix, and
$\bm A$ is an arbitrary $K$-dimensional matrix.
The result, given by Eq. (\ref{eq:term21}), indicates that the entire set of
unitary matrices that appear in the diagonalization
of all possible random matrix products $\bm \Xi^{\dagger} \Rr \bm \Xi$ 
coincides with the entire set of unitary matrices,
which guarantees that the matrix integration formula over the unitary matrix
is applicable only to the $\bm \Xi^{\dagger} \Rr \bm \Xi$
part of the entire cross-correlation matrix $\bm H^{\dagger} \bm H$.
This is because the multiplication of the matrix $\bm \Xi$ serves as 
a unitary transformation. Note that we cannot apply the same
argument to the entire cross-correlation matrix
$\bm H^{\dagger} \bm H =
\sqrt{\Rt} \bm \Xi^{\dagger} \Rr \bm \Xi \sqrt{\Rt}$
because the transmitter correlation matrix $\Rt$ breaks
rotational invariance, and more careful treatment is required, as described below.

By using the saddle point method
and assuming replica symmetry
as $q=Q_{ab}$ for $a \ne b$ and $q+\chi = Q_{aa}$,
we can evaluate the replicated partition function in Eq. (\ref{eq:term21})
after introducing auxiliary variables $\hat{q}+\hat{\chi}$ and $-2\hat{q}$
for the delta functions of the diagonal and off-diagonal matrix elements,
respectively,
\begin{multline}
\label{eq:Sn}
\frac{1}{K}\ln \Pi^{(n)}(\bm Q)
=\extr_{\widehat{q},\widehat{\chi}}\Biggl\{
{n}\chi\widehat{\chi}
+{(\widehat{\chi}-(n+1)\widehat{q})(\chi+(n+1)q)}
\\+\frac{1}{K}\ln
\lb(
\lb\{\prod_{a=0}^{n} \Tr_{\bm b^a} P(\bm b^a)\rb\}
\exp\lb[
-\widehat{\chi}\sum_{a=0}^n\bm b^{a\dag}\Rt\bm b^a
+{\widehat{q}}\sum_{a b}^n\bm b^{a\dag}{\Rt}\bm b^{b}
\rb]
\rb)
\Biggr\}.
\end{multline}
The saddle point condition for $q$ yields $\widehat{q}=\widehat{\chi}/(n+1)$. 
After performing Hubbard-Stratonovich transformation,
\begin{equation}
\exp\lb[
-\widehat{\chi}\sum_{a=0}^n\bm b^{a\dag}\Rt\bm b^a
+\frac{\widehat{\chi}}{n+1}\sum_{a  b}^n\bm b^{a\dag}{\Rt}\bm b^{b}
\rb] = \left( \frac{(n+1) \widehat{\chi}}{\pi} \right)^{K}
\int_{\Complex^K} d\bm r' 
 \exp\lb[ - \widehat{\chi} \sum_{a=0}^{n}
\lb|{\bm r'}-\sqrt{\Rt}\bm b^{a} \rb|^2 \rb]
,
\end{equation}
we have
\begin{eqnarray} 
\label{eq:decoupled}
\frac{1}{K}\ln \Pi^{(n)}(\bm Q)
&=& \extr_{\widehat{\chi}}\Biggl\{
{n}\chi\widehat{\chi}
+\frac{1}{K}\ln
\int_{\Complex^K} d{\bm r'}
 \lb\{
\Tr_{\bm b}P(\bm b) \left( \frac{\widehat{\chi}}{\pi} \right)^K
\exp\lb[
-\widehat{\chi}\lb|{\bm r'}-\sqrt{\Rt}\bm b\rb|^2
\rb]
\rb\}^{n+1}
+n\ln \left( \frac{\pi}{\widehat{\chi}} \right) + \ln(n+1)
\Biggr\}.
\nonumber \\
&&
\end{eqnarray} 
Combining this equation with the remainder of the replicated partition function and
noting that the matrix $\calbmP \bm Q$ has a single
zero eigenvalue and $n$-degenerate $\chi$ under
the replica-symmetric condition, we obtain the final expression of the
mutual information, as follows:
\begin{eqnarray}
\label{eq:FreeEnergyGeneral}
\overline{{\cal I}_{\bm H}}
&=&
\extr_{\chi,\widehat{\chi}}\lb\{
-G_{\bm\Xi^\dag\Rr\bm\Xi}\lb(-\frac{\chi}{\sigma^2}\rb)
-\frac{\partial}{\partial n}
\frac{1}{K}\ln \Pi^{(n)}(\bm Q)\bigg|_{n=0}
\rb\}
\nonumber \\
&=&
\extr_{\lambda}\lb\{
\widehat{G}_{\bm\Xi^\dag\Rr\bm\Xi}\lb(\lambda\rb)
+
I_{\Rt}\lb(\frac{{\lambda}}{\sigma^2}\rb)
\rb\},
\end{eqnarray}
where $\widehat{G}_{\bm\Xi^\dag\Rr\bm\Xi}\lb(\lambda\rb)$ is
the Legendre transform of $G_{\bm\Xi^\dag\Rr\bm\Xi}(\lambda)$,
$\widehat{G}_{\bm\Xi^\dag\Rr\bm\Xi}\lb(\lambda\rb)
\equiv \extr_{\chi}\lb\{\lambda\chi-G_{\bm\Xi^\dag\Rr\bm\Xi}(\chi) \rb\}$,
and 
$I_{\Rt}\lb(\chi\rb)$ is defined as
\begin{eqnarray}
\label{eq:FreeEnergyGeneral2}
I_{\Rt}\lb(\chi\rb) \!\!
&\equiv& \!\!
-\frac{1}{K}
\int_{\Complex^K} d \bm r'
\lb\{ \Tr_{\bm b}P(\bm b) \left( \frac{\chi}{\pi} \right)^{K}
\exp\lb[
-{\chi\lb|{\bm r'}-\sqrt{\Rt}\bm b\rb|^2}
\rb] \rb\}
\ln 
\lb\{
\Tr_{\bm b}P(\bm b) \left( \frac{\chi}{\pi} \right)^{K}
\exp\lb[
-{\chi\lb|{\bm r'}-\sqrt{\Rt}\bm b\rb|^2}
\rb]
\rb\} \nonumber \\
&& \hspace{12cm} -\ln \left( \frac{\pi}{\chi} \right)-1.
\end{eqnarray} 
Equation (\ref{eq:FreeEnergyGeneral}) is the primary result of the present paper. 
As we demonstrate in Section \ref{sec:exp}, it is convenient to use
$I_{\Rt}(\chi)$ for the discussion of the performance of the channel.

In the following, we consider three items.
First, Eq. (\ref{eq:FreeEnergyGeneral}) provides
a physical meaning for the performance analysis of the Kronecker channel. 
The term $I_{\Rt}$ of the right-hand side corresponds to the mutual information of
a channel after rescaling of $\bm r'$,
\begin{equation}
\label{eq:submodel0}
{\bm r'}= (\sqrt{\lambda}/\sigma) \sqrt{\Rt}\bm b^0+ \bm \eta',
\end{equation} 
(See Eq. (\ref{eq:mutualinfo}))
where $\bm \eta'$ is a $K$-dimensional normal complex Gaussian noise. 
Here, we let $\lambda$ be a random variable that obeys probability
distribution $P(\lambda)\simeq \exp\lb[K\widehat{G}_{\bm\Xi^\dag\Rr\bm\Xi}(\lambda)
\rb]$. Then, Eq. (\ref{eq:FreeEnergyGeneral}) means that, in the limit,
$K \rightarrow \infty$ $\overline{{\cal I}_{\bm H}}$
corresponds to the average of the exponential of the mutual information,
$\exp(K I_{\Rt} (\lambda/\sigma^2))$ over $\lambda$.
The extremization of Eq. (\ref{eq:FreeEnergyGeneral})
implies that the balance of the two $\lambda$-dependent functions
$\widehat{G}_{\bm \Xi^{\dagger} \Rr \bm \Xi} (\lambda)$
and $I_{\Rt} (\lambda/\sigma^2)$, which are dependent on
correlations among reception antennas and among transmit antennas, respectively,
is significant in the determination of $\overline{{\cal I}_{\bm H}}$.

Second, $I_{\Rt}(\chi)$ can be evaluated using the following approximation method. After performing unitary transformation of the matrix
$\Rt$ to $\bm U^\dag\Rt\bm U$, 
we take the average of $I_{\bm U^\dag\Rt\bm U}(\chi)$ over unitary matrix $\bm U$
(denoted by $\overline{I_{\bm U^{\dagger} \Rt \bm U} (\chi)}$ 
in the following) as in the case for the matrix $\bm \Xi\Rr\bm \Xi^{\dagger}$.
In a manner similar to the evaluation of ${\cal I}_{\bm H}$, we have
\begin{eqnarray}
\overline{I_{\bm U^\dag\Rt\bm U}(\chi)} &=& \extr_{\lambda}
\lb\{\widehat{G}_{\Rt}(\lambda)+I_{\bm I}(\lambda\chi) \rb\},
\end{eqnarray}
where $I_{\bm I}(\chi)$ is the mutual information of
Eq. (\ref{eq:FreeEnergyGeneral2}) after the substitution of $\bm R_{t}=\bm I$,
which can be decomposed to the mutual informations of multiple one-dimensional channels.

\def\Rnd{\bm R}
Third, if the correlation among transmission antennas is sufficiently
small, we can perform a perturbative expansion of $I_{\bm\Rt}$.
Let us consider the case in which the matrix $\Rt$ is expressed as 
$\Rt =\bm I+\rho\Rnd$ with a real small parameter $\rho$ 
and $K$-dimensional matrix $\Rnd$, the diagonal elements of which are all zero. 
After expansion, we have
\begin{eqnarray}
I_{\bm I+\rho\bm R}(\chi)
&=&
I_{\bm I}(\chi)
-\frac{(\rho \chi)^2}{2}
 \frac{\Tr(\bm R^2)}{K} \{I'_{\bm I}(\chi) \}^2
+\frac{(\rho \chi)^3}{3}
 \frac{\Tr(\bm R^3)}{K} \{I'_{\bm I}(\chi) \}^3
+\cdots,
\end{eqnarray}
where $I'_{\bm I}(\chi)$ is the derivative of
 $I_{\bm I}(\chi)$ with respect to $\chi$. 
Similarly, expanding the approximate mutual information
$\overline{I_{\bm I+\rho\bm U^\dag\bm R\bm U}(\chi)}$, 
we obtain the same result up to the third order of $\rho$. 
However, a discrepancy appears starting from the fourth-order coefficient.
In the case of QPSK modulation, this 
discrepancy is expressed as 
(see the Appendix)
\begin{eqnarray}
\label{eq:expansion4_diff}
I_{\bm I+\rho\bm R}(\chi)
-\overline{I_{\bm I+\rho\bm U^\dag\bm R\bm U}(\chi)}
&=&
-\frac{(\rho \chi)^4}{2}\frac{1}{K}
\sum_{i}\lb\{
(\bm R^2)_{ii}- \frac{\Tr(\bm R^2)}{K}
\rb\}^2
\lb\{-I_{\bm I}''(\chi)-I_{\bm I}'(\chi)^2\rb\}\lb\{I'_{\bm I}(\chi)\rb\}^2
\nonumber \\
&& \hspace{-25mm}
-\frac{(\rho \chi)^4}{4}\frac{1}{K} \left( \sum_{ij}
 \left\{ \Re (R_{ij})^4 + \Im (R_{ij})^4 \right\} \right)
\left( \lb\{-I_{\bm I}''(\chi)-I_{\bm I}'(\chi)^2\rb\}^2
 + \frac{C(\chi)^2}{6} \right) 
+O(\rho^5),
\end{eqnarray}
where $C(\chi)$ is a function that depends on $P(\bm b)$ as well as $\chi$.
This indicates that the approximate evaluation of mutual information
by matrix integration yields a good result
if the perturbation parameter $\rho$ is sufficiently small.
Under this condition, the evaluation using matrix integration
described in \cite{bib:THK2007} has an advantage in that it provides a good
approximate solution that is more convenient than the exact evaluation of
mutual information for the channel $\bm r=\bm H\bm b+\bm\eta$.

As described in the Appendix, we can prove that
$\lb\{-I_{\bm I}''(\chi)-I_{\bm I}'(\chi) ^2\rb\}\geq 0$,
and, accordingly, the right-hand side of Eq. (\ref{eq:expansion4_diff}) becomes 
non-positive for a wide class of $P(\bm b)$, including QPSK
modulation, which means that approximate evaluation by 
$\overline{I_{\bm I+\rho\bm U^\dag\bm R\bm U}(\chi)}$
gives an upper bound of $I_{\bm I+\rho\bm R}(\chi)$ 
up to the forth order of the correlation parameter $\rho$. 

%%%%%%%%%%%%%%%%%%%%%%%%%%%%%%%%%%%%%%%%%%%%%%%%%%%%%%%%%%%%%%%%%%%
%%%%%%%%%%%%%%%%%%%%%%%%%%%%%%%%%%%%%%%%%%%%%%%%%%%%%%%%%%%%%%%%%%%

\section{Demodulation Algorithm}
\label{sec:algo}

For practical communication, it is also significant 
to construct a computationally feasible demodulation algorithm.
For inference of original signal $\bm b$ 
from received signal $\bm r$ and channel matrix $\bm H$, 
it is necessary to evaluate the following quantity:
\begin{equation}
\label{eq:mdef}
\bm m =\sum_{\bm b\in{\cal B}^K} \bm{b} P(\bm b|\bm r,\bm H).
\end{equation} 
However, it is computationally 
difficult to numerically evaluate $\bm m$ from this expression.
Key to the practical solution of this problem is
the use of the Gibbs free energy for the communication channel:
\begin{equation}
\label{eq:Gibseval}
\Phi(\bm m)=\extr_{\bm h}\lb\{
-\ln\Tr_{\bm b}P(\bm b|\bm r,\bm H) \exp[\Re(\bm h^\dag (\bm b - \bm m) )]
\rb\},
\end{equation}
and the quantity $\bm m$ can be estimated as the argument of the
extremized Gibbs free energy. 
Substituting 
$P(\bm b|\bm r,\bm H)
=P(\bm b)\exp\lb[-{|\bm r-\bm H\bm b|^2}/{\sigma^2}\rb]/Z$
with $Z$ being the normalization
and $\bm H=\sqrt{\Rr} \bm \Xi \sqrt{\Rt}$ we have
\begin{equation}
\label{eq:ggg0}
\Phi(\bm m)=\extr_{\bm h}\lb\{
\frac{|\bm r-\bm H\bm m|^2}{\sigma^2}
-\ln\Tr_{\bm b}\lb( P(\bm b)
\exp\lb[-\frac{|\sqrt{\Rr} \bm \Xi \sqrt{\Rt} (\bm b - \bm m)|^2}{\sigma^2}
+\Re(\bm h^\dag (\bm b - \bm m))\rb]
\rb)
\rb\}
+\ln Z.
\end{equation} 
Note that the extremization argument $\bm h$ is shifted 
as $\bm h + 2 (\bm H^{\dagger} \bm r- \bm H^{\dagger} \bm H \bm m
) / \sigma^2 \rightarrow \bm h $.
Although this distribution of the vector $\sqrt{\Rt}(\bm b - \bm m)$
is not isotropic, but rather is biased by the matrix $\sqrt{\Rt}$, 
the multiplication by the rectangular random matrix $\bm\Xi$
ensures the following approximation under the constraints 
$\chi=|\sqrt{\Rt} (\bm b - \bm m)|^2/K$
and $\kappa=\Re(\bm h^\dag(\bm b - \bm m))/K$,
where we introduce the auxiliary variables $\chi$ and $\kappa$, as follows:
\begin{eqnarray}
\!\!\! &&\Tr_{ \bm b}
P(\bm b)
\exp\lb[-{|\sqrt{\Rr} \bm \Xi \sqrt{\Rt} (\bm b - \bm m)|^2}/{\sigma^2}
+\Re(\bm h^\dag (\bm b -\bm m))\rb]
\nonumber \\
\!\!\! &\simeq& \!\!\!\! \int d\chi \int d \kappa
\exp\lb[K\lb\{G_{\bm \Xi^{\dagger} \Rr \bm \Xi}
\lb(-{\chi}/{\sigma^2}\rb)+\kappa\rb\}\rb] 
\Tr_{\bm b}\lb\{P(\bm b)\,
\delta\lb(|\sqrt{\Rt}(\bm b - \bm m)|^2-K\chi\rb)
\delta\lb(\Re(\bm h^\dag(\bm b - \bm m))-K\kappa\rb)\rb\},
\end{eqnarray}
where $G_{\bm \Xi^{\dagger} \Rr \bm \Xi}(x)$ is given 
by Eq. (\ref{eq:Gfunc}). Saddle point evaluation yields the following approximate
expression of the Gibbs free energy:
\begin{eqnarray}
\label{eq:ApGibbs}
\Phi(\bm m)&\simeq&
\extr_{\chi, \widehat{\chi}}\lb\{
\frac{|\bm r-\bm H\bm m|^2}{\sigma^2} 
-K G_{\bm \Xi^{\dagger} \Rr \bm \Xi}
\lb(-\frac{\chi}{\sigma^2} \rb) 
-K\widehat{\chi}\chi
+\Phi_{\mathrm{t}}(\bm m;\widehat{\chi})
\rb\}+\ln Z,
\nonumber \\
\label{eq:ApGibbs2}
{\rm where \ \ }
\Phi_{\mathrm{t}}(\bm m;\widehat{\chi})&=&
\extr_{\bm h}\lb\{
-\ln\Tr_{\bm b} \lb(\exp\lb[
-{\widehat{\chi}} | \sqrt{\Rt} (\bm b - \bm m)|^2
+\Re(\bm h^\dag (\bm b-\bm m))
\rb]\rb)\rb\}.
\end{eqnarray}

From the Gibbs free energy we obtain a set of equations for estimating $\bm m$,
and using these equations, we construct the demodulation
algorithm, or the method for finding the minimization argument $\bm m$
for the Gibbs free energy. In the following, we summarize the procedure for 
the minimization of $\Phi(\bm m)$. 
\begin{itemize}
\item (Step 0) Initialize variables as 
$\chi^{(0)} = 1, \bm m^{(0)} = {\bm 0}, \bm h^{(0)} = {\bm 0}$ 
for step $t=0$, and set the number of steps as $t=1$.
\item (Step 1) For the $t$-th step, update $\widehat{\chi}$ and $\bm h$ as 
\begin{eqnarray}
\widehat{\chi}^{(t)}&=&
\frac{1}{\sigma^2}G'\lb(-\frac{\chi^{(t-1)}}{\sigma^2}\rb),
\nonumber \\
\bm h^{(t)}&=&
\bm h^{(t-1)}+ \sigma^2 \widehat{\chi}^{(t)}
\lb( 2 \ \frac{ \bm H^\dag\bm r-\bm H^\dag\bm H\bm m^{(t-1)}}{\sigma^2}
-\bm h^{(t-1)} \rb).
\nonumber
\end{eqnarray}
\item (Step 2) Update $\bm m$ as 
\begin{eqnarray} 
\bm m^{(t)}&=& 
\langle \bm b \rangle_{t},
\nonumber \\
\chi^{(t)}&=&
\frac{1}{K}\lb\{\langle
\bm b^\dag\Rt\bm b \rangle_{t}
-\bm m^{(t)\dag}\Rt\bm m^{(t)}\rb\},
\nonumber
\end{eqnarray}
where $\langle \cdot \rangle_{t}$ denotes
 the expectation  
\[
\langle f(\bm b) \rangle_{t} \equiv
\frac{\Tr_{\bm b}P(\bm{b})\exp\lb[
-{\widehat{\chi}^{(t)}}
|\sqrt{\Rt} (\bm b - \bm m) |^2 +\Re(\bm h^\dag\bm b) \rb] f(\bm b) }
{\Tr_{\bm b}P(\bm{b})\exp\lb[
-{\widehat{\chi}^{(t)}} |\sqrt{\Rt} (\bm b - \bm m) |^2
 + \Re(\bm h^\dag\bm b) \rb]}.
\]

\item (Step 3) Update the number of recursion steps $t\rightarrow t+1$.
Return to Step 1 unless these variables converge, otherwise stop. 
\end{itemize}
After termination of the above procedure, the transmit signal is estimated
as $\widehat{\bm b}=\argmin_{\bm b}\lb\{|\bm b-\bm m^{(t)}|\rb\}$. 

The computational cost of Step 1 is $O(KL)$ and is sufficiently small.
In general, the cost of Step 2 is not so small.
However, we can reduce the cost of Step 2 for the special forms
of matrix $\Rt$.
For instance, when $\Rt$ is a matrix of tridiagonal form, as considered
in the following, Step 2 can be executed using 
the transfer matrix method. 
In such a case, the cost is $O(K)$, which is smaller than the cost of Step 1.

%%%%%%%%%%%%%%%%%%%%%%%%%%%%%%%%%%%%%%%%%%%%%%%%%%%%%%%%%%%%%%%%%%%
%%%%%%%%%%%%%%%%%%%%%%%%%%%%%%%%%%%%%%%%%%%%%%%%%%%%%%%%%%%%%%%%%%%

\section{Numerical Experiment} 
\label{sec:exp}

As a simple but nontrivial example, we performed numerical experiments for
using Kronecker channel model $\bm H=\sqrt{\Rr}\bm\Xi\sqrt{\Rt}$, the
correlation matrices $\Rr$ and $\Rt$ of which are identity and tridiagonal matrices, respectively. 
More precisely, we consider the case in which matrix $\Rt$ has
nonzero elements only for adjacent antennas, modeled by $\Rt=\bm I+\rho\bm R$,
where the matrix $R_{kk'} = R_{k'k} \equiv \delta_{k(k'-1)} (k \le k')$
 and $\rho$ is a parameter that represents the strength of
correlations.
For simplicity, we analyze the real channel, and, accordingly, all
variables are set to be real.
 Here, $\bm\Xi$ represents the $L\times K$-dimensional
i.i.d. Gaussian random matrix ${\cal N}(0, 1/L)^{KL}$, 
$\bm b=(b_k)\in\Real^K$ is a BPSK-modulated transmit signal
($b_k\in\{\pm 1\}$), and
$\bm\eta\in\Real^L$ is a normalized real Gaussian-distributed
random vector ${\cal N}(0, 1)$. 
The formulation so far for the complex channel can be reconstructed without
difficultly for the real channel just by the replacement of 
unitary matrix $\bm U$ with orthogonal matrix $\bm O$.

%%%%%%%%%%%%%%%%%%%%%%%%%%%%%%%%%%%%%%%%%%%%%%%%%%%%%%%%%%%%%%%%%%%

\subsection{Random-field Ising chain} 

Before the analysis of the entire Kronecker model, 
let us evaluate three pieces of mutual information, namely,  
$I_{1}(\chi)$, $I_{\bm I+\rho\bm O^{T} \bm R\bm O}(\chi)$
and $I_{\bm I+\rho\bm R}(\chi)$ \cite{bib:diffrealcomplex}, that
appear in the expression of the mutual information of the entire system
${\cal I}_{\bm H}$, as described in Section \ref{sec:analysis}. 
For the BPSK modulation, 
the mutual information of the one-dimensional channel $I_{1}(\chi)$ is given by 
\begin{eqnarray}
I_{1}(\chi)&=& \chi - \frac{1}{2} \int Dz \ln\lb(\cosh(\chi+\sqrt{\chi}z)\rb),
\end{eqnarray} 
where $Dz \equiv \exp (-z^2/2)/ \sqrt{2 \pi}$.
For mutual information
$\overline{I_{\bm I + \rho \bm O^{T} \bm R \bm O}(\chi)}$,
substitution of the tridiagonal form of $\bm R$ yields 
\begin{eqnarray}
\overline{I_{\bm I+\rho\bm O^{T} \bm R\bm O}(\chi)}
&=&
\extr_{\lambda} \lb\{
\widehat{G}_{\bm I+\rho\bm O^{T} \bm R\bm O}(\lambda) + I_{1}(\lambda\chi)
\rb\}.
\end{eqnarray}
Here, $\widehat{G}_{\bm I+\rho\bm O^{T} \bm R\bm O}(\lambda)
=-(1/2)\ln\lb(1-(\lambda-1)^2/4\rho^2\rb)$,
which is evaluated using the Stieltjes inversion formula
for the function $G(\chi)$ (see Reference {\cite{bib:TUK2006}) and
the relation $\rho_{\Rt}(\lambda) =
 \lim_{\epsilon \rightarrow 0} (\pi N)^{-1} \partial_{\lambda} {\rm Im}
 \ln \det (\Rt - (\lambda - i \epsilon) \bm I) $. 
As described earlier, the discrepancy
between $I_{\bm I+\rho\bm R}(\chi)$ and
$\overline{I_{\bm I+\rho\bm O^{T} \bm R\bm O}(\chi)}$ appears
starting from the fourth-order term,
which is expressed as the tridiagonal form of $\bm R$, as follows:
\begin{eqnarray}
I_{\bm I+\rho\bm R}(\chi)
&=&
\overline{I_{\bm I+\rho\bm O^{T} \bm R\bm O}(\chi)}
- \frac{(\rho\chi)^4}{4}
\left[ \lb\{-2I''_1(\chi)-(2I'_1(\chi))^2\rb\}^2 +\frac{\widehat{C}(\chi)^2}{6} \right]
+O(\rho^5),
\end{eqnarray} 
where
\begin{equation}
\widehat{C}(\chi) \equiv -2 \int Dz \left\{ 1 - \tanh^2(\chi + \sqrt{\chi} z) \right\}
\left\{ 1 - 3 \tanh^2 (\chi + \sqrt{\chi} z) \right\}.
\end{equation}

For the tridiagonal form of $\bm R$,
we can evaluate $I_{\bm I+\rho\bm R}(\chi)$ exactly 
using the transfer matrix method for the Ising chain in random fields.
In order to demonstrate how this is accomplished, we transform the mutual information as follows:
\begin{eqnarray}
\label{eq:Isingchain}
I_{\bm I+\rho\bm R}(\chi)
&=& 
-\frac{1}{K}
\int_{\Real^K} d \bm r'
\lb\{ \Tr_{\bm b} \frac{1}{2^K}
\lb(\frac{\chi}{2 \pi}\rb)^{K/2}
 \exp\lb[ -{\frac{\chi}{2}
\lb|{\bm r'}- \sqrt{\bm I + \rho \bm R}\ \bm b\rb|^2}
\rb] \rb\} \nonumber \\
&& \hspace{2cm} \times \ln \lb\{
\Tr_{\bm b}  \frac{1}{2^K} 
\lb(\frac{\chi}{2 \pi}\rb)^{K/2}
 \exp\lb[
-{\frac{\chi}{2}\lb|{\bm r'}- \sqrt{\bm I + \rho \bm R}\ \bm b\rb|^2}
\rb] \rb\} 
- \frac{1}{2} \ln \left( \frac{2 \pi}{\chi} \right)
- \frac{1}{2} \nonumber \\
&\simeq&
-\frac{1}{2^K K} \int_{\Real^K} D\bm\eta \Tr_{\bar{\bm b}}
\ln\lb\{\Tr_{\bm b}
\prod_{k=1}^{K-1}\phi(b_k, b_{k+1}|\bar{b}_{k}, \bar{b}_{k+1}, \eta_k)
\rb\}
\nonumber \\
&=&
-\frac{1}{2^K K} \int_{\Real^K} D\bm\eta \Tr_{\bar{\bm \tau}}
\ln\lb\{\Tr_{\bm\tau}
\prod_{k=1}^{K-1}\phi(\tau_k, \tau_{k+1}|\bar{\tau}_k, \eta_k)
\rb\},
\end{eqnarray}
where a trivial constant is neglected in the second and the third lines.
Note that gauge transformation as $\bm b\to\bm\tau$
and $\bar{\bm b}\to\bar{\bm \tau}$,
where $\tau_k=b_k \bar{b}_k$ and
$\bar{\tau}_{k}=\bar{b}_{k+1}\bar{b}_{k}$, 
and redefinition of $\bm \eta$ are used for simplification of the expression.
Matrix elements $\phi(b_k, b_{k+1}|\bar{b}_{k}, \bar{b}_{k+1}, \eta_k)$
and $\phi(\tau_k, \tau_{k+1}|\bar{\tau}_{k},\eta_k)$ are defined as follows:
\begin{eqnarray}
\label{eq:RIfactor}
\phi(b_k, b_{k+1}|\bar{b}_{k}, \bar{b}_{k+1}, \eta_k)
&=&\frac{1}{2}
\exp\lb[
-\frac{\chi}{2}\lb|l_0 (b_k-\bar{b}_k)
+ l_1(b_{k+1}-\bar{b}_{k+1})\rb|^2
+\sqrt\chi\eta_k\lb(
 l_0 b_k+ l_1 b_{k+1}\rb)
\rb],
\nonumber \\
\phi(\tau_k, \tau_{k+1}|\bar{\tau}_{k},\eta_k)
&=&\frac{1}{2}	\exp\lb[
-\frac{\chi}{2}\lb|l_0 (\tau_k-1)+ l_1
\bar{\tau}_k(\tau_{k+1}-1)\rb|^2
+\sqrt\chi\eta_k\lb(
 l_0 \tau_{k} + l_1 \bar{\tau}_k \tau_{k+1} \rb)
\rb],
\end{eqnarray} 
where $l_0$ and $l_1$ are real constants
that are obtained by Cholesky decomposition, i.e.,
$\bm I+\rho \bm R = \bm \Lambda \bm \Lambda^{T}$, where
$\Lambda_{kk}=l_0, \Lambda_{(k+1) k} = l_1 $, and zero otherwise.
These constants satisfy $l_0^2+l_1^2=1$, $l_0l_1=\rho$, and $l_0\geq l_1$. 
The matrix element of Eq. (\ref{eq:RIfactor}) corresponds to the Boltzmann weight
of the Ising chain coupled with bimodal ($\bar{\bm \tau}$)
and Gaussian ($\bm \eta$) random fields,
and consequently the bit error rate (BER) for the demodulation
$\widehat{b}_k=\argmax_{b_k}\lb\{\Tr_{\bm b\backslash b_k}
P(\bm b|\bm r)\rb\}$
is evaluated analytically for the random field Ising chain.

Several methods of analysis have been developed for handling
the random field Ising chain
\cite{bib:ising_Derrida, bib:ising_Weigt}, and in the present study, we
use the technique of belief propagation,
which is equivalent to the transfer matrix method.
After parameterization of the belief from site $k$ to site $k+1$ by
cavity field $h_{\rightarrow k+1}$ as $\mu(\tau_{k+1}) = e^{h_{\rightarrow k+1}
\tau_{k+1}} / 2 \cosh(h_{\rightarrow k+1})$, the propagation process from 
$h_{\rightarrow k}$ to $h_{\rightarrow k+1}$ is written as
\begin{eqnarray}
h_{\rightarrow k+1} &=&
\arctanh \left( \sum_{\tau_{k+1}} \mu (\tau_{k+1}) \tau_{k+1} \right)
= \arctanh \lb(
\frac{ \sum_{\tau_k  \tau_{k+1}} \phi(\tau_k,\tau_{k+1} |
\bar{\tau}_k \eta_k ) \exp( h_{\rightarrow k} \tau_k ) \tau_{k+1} }
{\sum_{\tau_k  \tau_{k+1}} \phi(\tau_k, \tau_{k+1} |
\bar{\tau}_k \eta_k ) \exp( h_{\rightarrow k}  \tau_{k}) }
\rb)
\nonumber \\
&=& \chi l_1^2 + \chi \rho \bar{\tau_{k}}
+ \sqrt{\chi} l_1 \bar{\tau}_k \eta_k - \bar{\tau}_k 
\arctanh (\tanh(\rho \chi) \tanh(h_{\rightarrow k} 
+ \chi l_0^2 + \chi \rho \bar{\tau}_k + \sqrt{\chi} l_0 \eta_k ) ).
\end{eqnarray}
Similarly, for the opposite direction, denoted by $h_{k \leftarrow}$,
\begin{eqnarray}
h_{k \leftarrow }
&=& \arctanh \lb(
\frac{ \sum_{\tau_k  \tau_{k+1}} \phi(\tau_k,\tau_{k+1} |
\bar{\tau}_k \eta_k ) \exp( h_{k+1 \leftarrow} \tau_{k+1} ) \tau_k }
{\sum_{\tau_k  \tau_{k+1}} \phi(\tau_k, \tau_{k+1} |
\bar{\tau}_k \eta_k ) \exp( h_{k+1 \leftarrow}  \tau_{k+1}) }
\rb)
\nonumber \\
&=& \chi l_0^2 + \chi \rho \bar{\tau_{k}} + \sqrt{\chi} l_0
\eta_k - \bar{\tau}_k 
\arctanh (\tanh(\rho \chi) \tanh(h_{k+1 \leftarrow} 
+ \chi l_1^2 + \chi \rho \bar{\tau}_k + \sqrt{\chi}
l_1 \bar{\tau}_k \eta_k ) ).
\end{eqnarray}
The stationary distributions of beliefs for the numerically increasing and
decreasing directions, denoted by ${\pi}_+$ and ${\pi}_-$, respectively,
satisfy the following conditions: 
\begin{eqnarray}
{\pi_+(h_{\rightarrow k+1})}&=&
\int{\pi_+(h_{\rightarrow k})d{h}_{\rightarrow k}}
\int D\eta
\sum_{\bar{\tau}=\pm1}\frac{1}{2} 
\delta\Biggl(
h_{\rightarrow k+1}-
({\chi} l_1^2+\chi \rho \bar{\tau}+\sqrt{{\chi}}l_1\bar{\tau}\eta)
\nonumber \\ && 
\hspace{20mm}
+\bar{\tau}\,\arctanh\lb(\tanh(\rho\chi)
\tanh(h_{\rightarrow k}+
{\chi}l_0^2+\chi \rho \bar{\tau}+\sqrt{\chi}l_0\eta)
\rb)
\Biggr),
\\ \nonumber
{\pi_-(h_{k \leftarrow})}&=&
\int{\pi_-(h_{k+1 \leftarrow})d{h}_{k+1 \leftarrow}}
\int D\eta
\sum_{\bar{\tau}=\pm1}\frac{1}{2} 
\delta\Biggl(
h_{k \leftarrow} -
({\chi}l_0^2+\chi \rho \bar{\tau}+\sqrt{{\chi}}l_0 \eta)
\\ && 
\hspace{20mm}
+\bar{\tau}\,\arctanh\lb(\tanh(\rho\chi)
\tanh(h_{k+1 \leftarrow}
+\chi l_1^2+\chi \rho \bar{\tau}+\sqrt{{\chi}}l_1 \bar{\tau} \eta)
\rb)
\Biggr),
\end{eqnarray} 
where $\int \pi_{+} (h_{\rightarrow}) d h_{\rightarrow}
= \int \pi_{-} (h_{\leftarrow}) d h_{\leftarrow} =1$.
The functions $\pi_+ (h_{\rightarrow k+1})$ and $\pi_- (h_{k \leftarrow})$
can be obtained numerically by the Monte Carlo method for a one-dimensional system.
The bit error rate $P_b$ is represented by
$P_b=\int {\pi}_+(h_{\rightarrow})dh_{\rightarrow}
\int{\pi}_- (h_{\leftarrow}) 
dh_{\leftarrow} \,
\lb(1-\mathrm{sgn}(h_{\rightarrow} +h_{\leftarrow} )\rb)/2$. 

\begin{figure}
\begin{picture}(180,165)
\put(-135,150){(a)}
\put(-130,0){\includegraphics[width=0.40\textwidth]{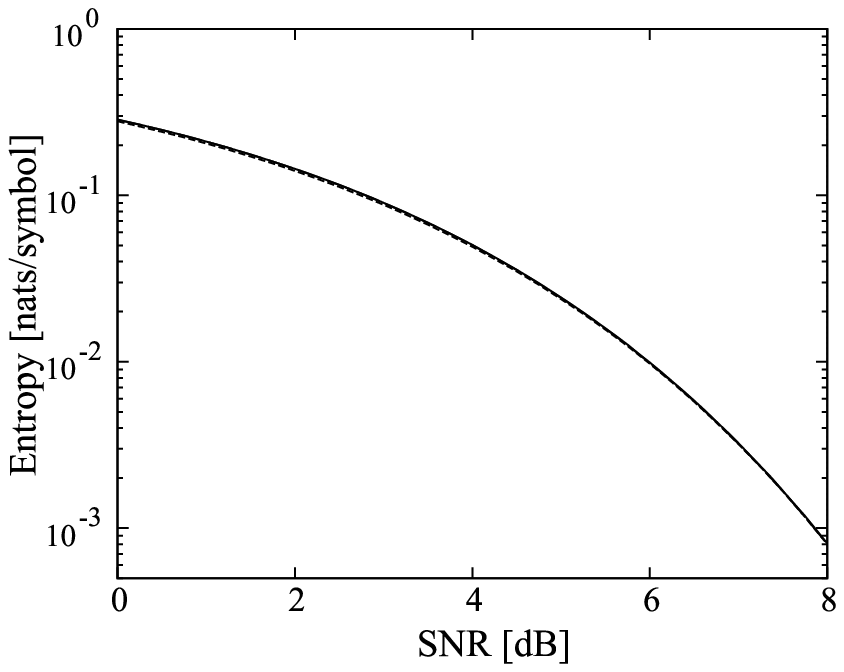}}
\put(95,150){(b)}
\put(100,0){\includegraphics[width=0.40\textwidth]{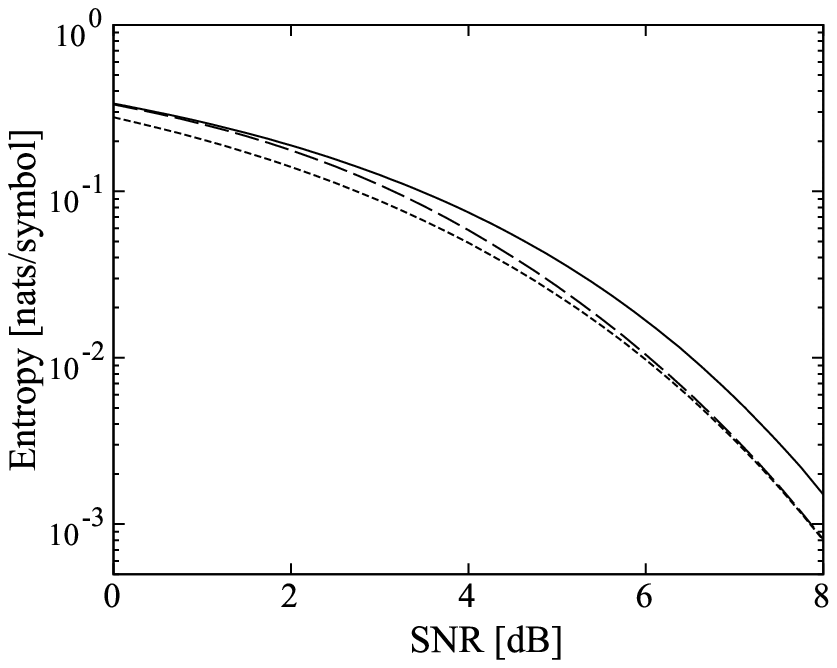}}
\end{picture}
\caption{Conditional entropy $h_{\bm I+\rho\bm R}(\bm b|\bm r)$
 for chain-like system (1) ((a) $\rho=0.2$, (b) $\rho=0.5$).
 The solid lines show the results obtained by exact analysis,
the broken lines show the results obtained by matrix integration, and the dotted lines show the results obtained by i.i.d. channel.} 
\label{fig:SNRvsI}
\begin{picture}(180,165)
\put(-135,145){(a)}
\put(-130,0){\includegraphics[width=0.40\textwidth]{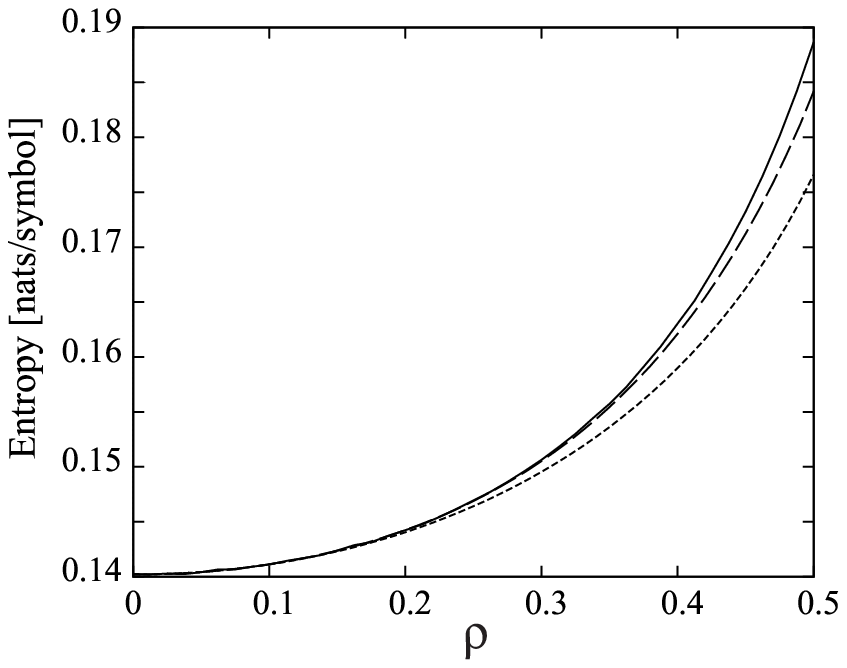}}
\put(95,145){(b)}
\put(100,0){\includegraphics[width=0.40\textwidth]{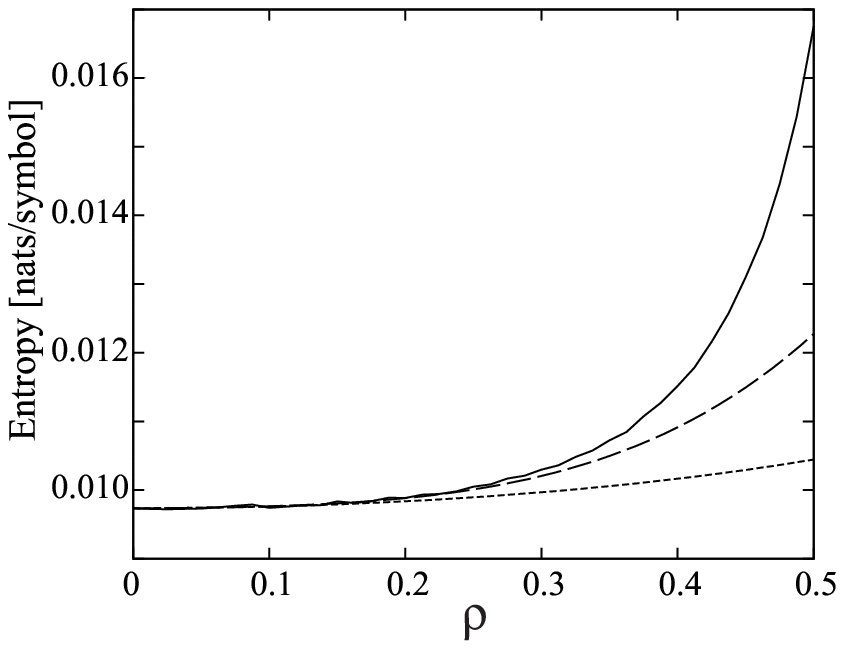}}
\end{picture}
\caption{Conditional entropy $h_{\bm I + \rho \bm R}(\bm b|\bm r)$
 for chain-like system (2). BER vs. correlation parameter $\rho$ 
 ((a) SNR = 2 dB, (b) SNR = 6 dB).
  The solid lines show the results obtained by exact analysis, the broken lines show the results obtained by matrix integration with correction from the fourth order, and the dotted lines show the results obtained by matrix
 integration without correction.} 
\label{fig:RhovsI}
\end{figure}
In order to investigate the performance of communication channels, 
conditional entropy $h(\bm b|\bm r)=h(\bm b)-{\cal I}_{\bm H}$,
where $h(\bm b)$ denotes entropy, 
is a favorable measure, because $h(\bm b|\bm r)$ decreases to zero
under smaller noise power.
Figures \ref{fig:SNRvsI} and \ref{fig:RhovsI}
show the exact conditional entropy
$h_{\bm I+\rho\bm R}(\bm b|\bm r)$
estimated by the Monte Carlo method, the approximate entropy
$\overline{h_{\bm I+\rho\bm O^{T} \bm R\bm O}(\bm b|\bm r)}$
obtained by matrix integration,
and the entropy of the i.i.d. channel $h_{\bm I}(\bm b|\bm r)$.
In both graphs, the entropy obtained by matrix integration, i.e.,
$\overline{h_{\bm I+\rho\bm O^{T} \bm R\bm O}(\bm b|\bm r)}$,
does not exceed the entropy obtained by exact evaluation 
$h_{\bm I+\rho\bm R}(\bm b|\bm r)$,
which implies the inequality $I_{\bm I+\rho \bm R}(\bm b, \bm r)\leq
\overline{I_{\bm I+\rho\bm O^{T} \bm R\bm O}(\bm b, \bm r)}$, which is
given up to the fourth order of perturbation in Section \ref{sec:analysis}.
The analysis indicates that the deviation of the approximate result from the exact result
depends on the signal-to-noise ratio (SNR) and the correlation parameter $\rho$.
For a small SNR and small correlation (= small $\rho$)
the deviation is small, which means that the entropy,
$\overline{h_{\bm I+\rho\bm O^{T} \bm R\bm O}(\bm b|\bm r)}$
obtained by matrix integration gives a good approximation of the exact entropy,
$h_{\bm I+\rho\bm R}(\bm b|\bm r)$,
while the deviation becomes greater in the case of a large SNR or large correlation.

As we have discussed earlier,
the approximate evaluation with matrix integration
is useful because this simplifies the analysis.
However, as shown by the numerical results for conditional entropy,
this method is only valid when the correlation is small.

%%%%%%%%%%%%%%%%%%%%%%%%%%%%%%%%%%%%%%%%%%%%%%%%%%%%%%%%%%%%%%%%%%%

\subsection{Kronecker channel} 

Next, we examine whether the proposed demodulation algorithm is
practical. In Figs. \ref{fig:Kronecker1} and \ref{fig:Kronecker2},  
the results of demodulation for the Kronecker channels are depicted.
These figures show two BER curves, namely,
the BER obtained by exact analysis of the model $\Rt= \bm I + \rho \bm R$
with tridiagonal $\bm R$, as described in the previous subsection, 
and the BER obtained by the correlation matrix multiplied
by an arbitrary orthogonal matrix,
$\Rt = \bm I + \rho \bm O^{T} \bm R \bm O$
and averaged over the orthogonal matrix $\bm O$. For the latter model,
matrix integration analysis 
can be applied due to multiplication of the orthogonal matrix.
We have proposed an appropriate demodulation algorithm for each evaluation.
For the former model, without the orthogonal matrix multiplication,
we can use the belief propagation algorithm proposed in the previous 
subsection. For the latter model, with orthogonal matrix multiplication,
the demodulation algorithm we proposed in \cite{bib:TUK2006,bib:THK2007}
based on matrix integration and the Thouless-Anderson-Palmer
method \cite{bib:TAP,bib:OpperWinther1,bib:OpperWinther2} is applicable.

The results of demodulation for each case
show good agreement with the results obtained by replica analysis.
For large noise power and large $\rho$
(Figs. \ref{fig:Kronecker1}(b) and \ref{fig:Kronecker2}(b)), 
the BER of the model without orthogonal-matrix multiplication
becomes larger than that with orthogonal-matrix multiplication,
which reflects the discrepancy of mutual information from higher-order values of $\rho$, as mentioned in Section \ref{sec:analysis}.
Therefore, in designing the demodulation algorithm for
such correlated channels, appropriate treatment of the correlation matrix
should be taken into consideration. We also examined the convergence speed of the algorithm. The proposed algorithm in the previous subsection requires the $O(K^2)$ matrix operation and the $O(K)$ belief propagation process in each step, and convergence of this algorithm requires dozens of iterations. Therefore, we conclude that this algorithm is computationally feasible. 

\begin{figure}
\begin{picture}(180,165)
\put(-145,150){(a)}
\put(-140,0){\includegraphics[width=0.40\textwidth]{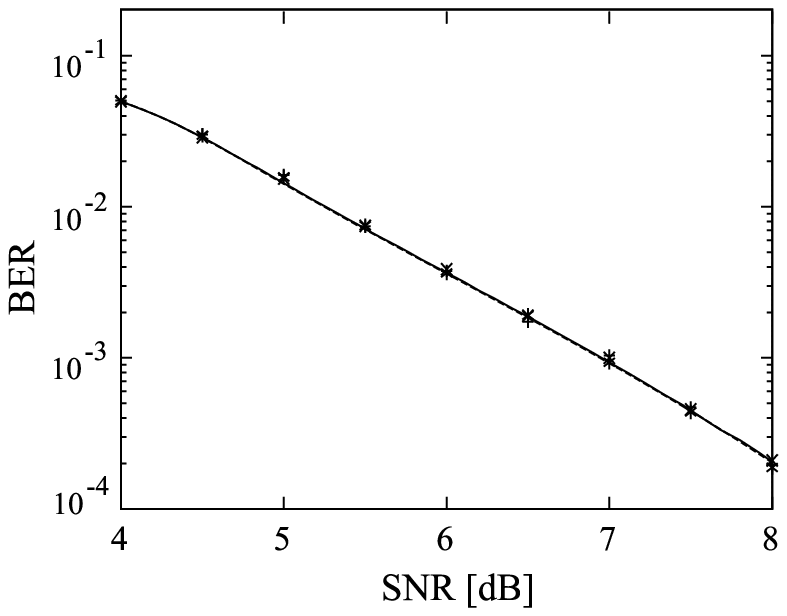}}
\put(85,150){(b)}
\put(90,0){\includegraphics[width=0.40\textwidth]{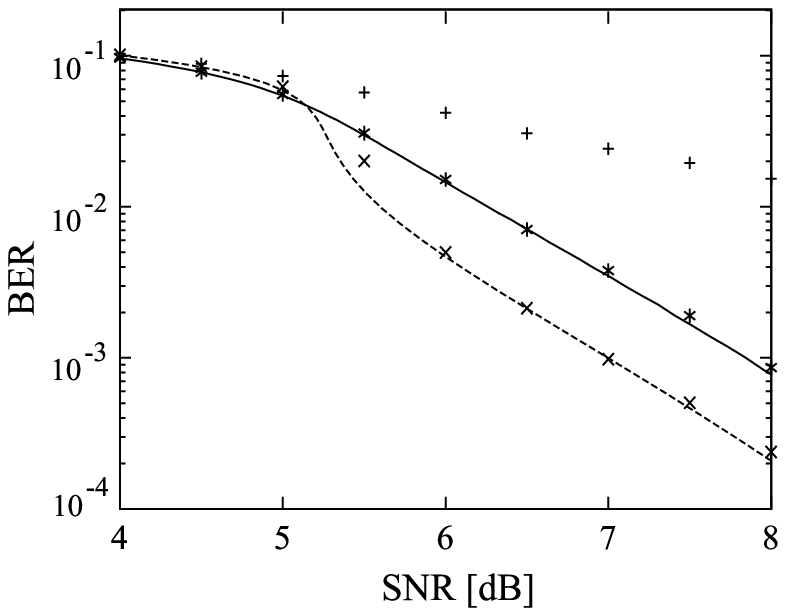}}
\end{picture}
\caption{Performance by the replica analysis and the result of
demodulation experiment (1). BER vs. SNR ((a) $\rho=0.2$, (b) $\rho=0.5$).
We set the parameters $K = 4,400$ and $L = 4,000$ and take the average of the results over 128 samples
with various input signals $\bm b$, matrices $\bm \Xi$, and noises $\bm \eta$.
We also varied the orthogonal matrix $\bm O$ for the model
with orthogonal-matrix multiplication.
The lines depict the results of the replica analysis for two models, the
chain-like model $\Rt = \bm I + \rho \bm R$ with
tridiagonal $\bm R$ (solid)
and the model with orthogonal-matrix multiplication 
$\Rt = \bm I + \rho \bm O^{T} \bm R \bm O$ (dotted). 
The symbols in the figure denote 
the results of demodulations, namely, demodulation for the chain-like model ($*$),
demodulation for the model with orthogonal-matrix multiplication ($\times$), 
inappropriate choice of the demodulation algorithm, i.e., the demodulation algorithm
for the orthogonal-matrix multiplication model applied to the chain-like model
($+$).}
\label{fig:Kronecker1}
\begin{picture}(180,165)
\put(-135,140){(a)}
\put(-130,-10){\includegraphics[width=0.40\textwidth]{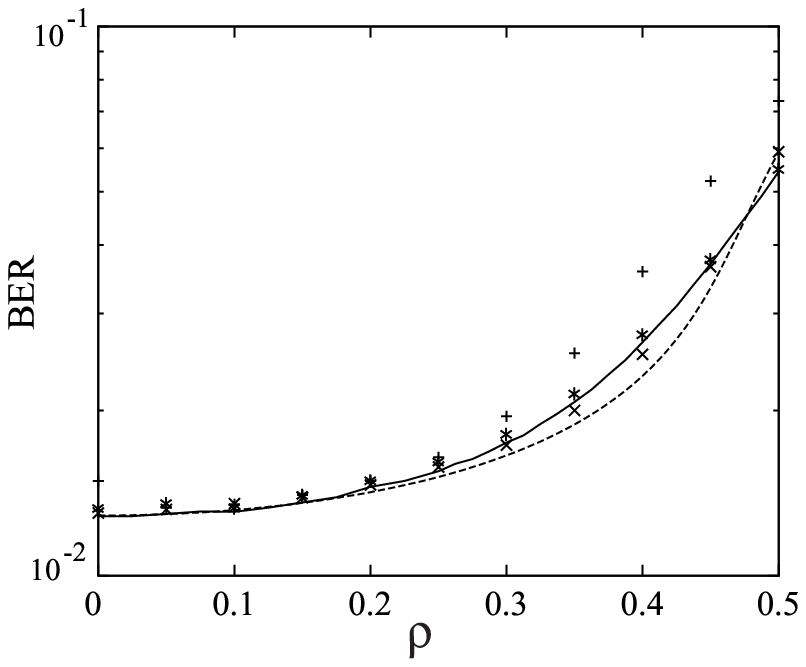}}
\put(95,140){(b)}
\put(100,-10){\includegraphics[width=0.40\textwidth]{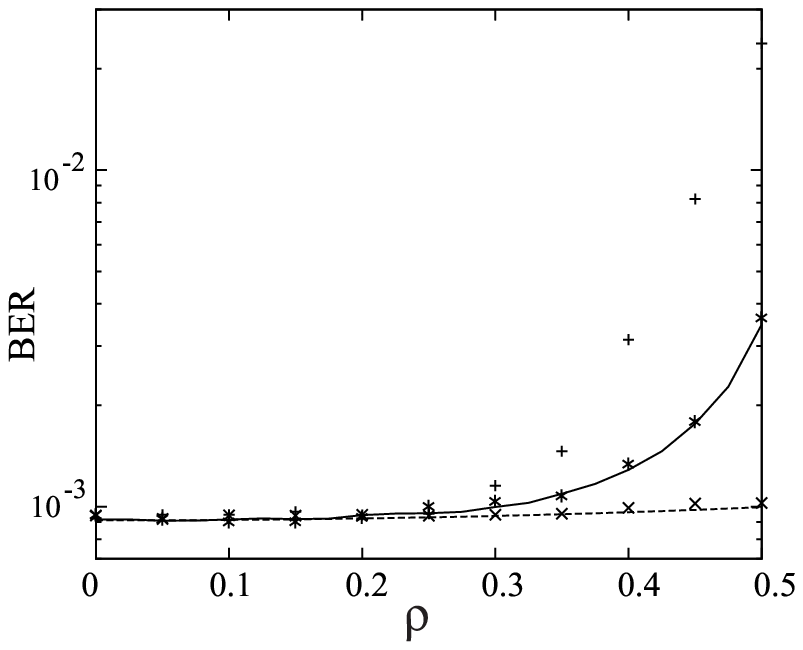}}
\end{picture}
\caption{Performance by the replica analysis and the results of
demodulation experiment (2). BER vs. correlation parameter $\rho$
((a) SNR = 5 dB, (b) SNR = 7 dB).
We set the parameter $K = 4,400$ and $L = 4,000$ and take the average over 1,024
samples for various values of $\bm b, \bm \Xi, \bm \eta$, and $\bm O$.
The lines depict the results of the replica analysis for two models, the
chain-like model $\Rt = \bm I + \rho \bm R$ with
tridiagonal $\bm R$ (solid)
and the model with orthogonal-matrix multiplication 
$\Rt = \bm I + \rho \bm O^{T} \bm R \bm O$ (dotted).
The symbols in the figure denote the results of demodulations, namely, demodulation for the chain-like model ($*$), demodulation for the model with orthogonal-matrix multiplication ($\times$), 
inappropriate choice of the demodulation algorithm, i.e., the demodulation algorithm
for the orthogonal-matrix multiplication model applied to the chain-like model
($+$).} 
\label{fig:Kronecker2}
\end{figure}

%%%%%%%%%%%%%%%%%%%%%%%%%%%%%%%%%%%%%%%%%%%%%%%%%%%%%%%%%%%%%%%%%%%
%%%%%%%%%%%%%%%%%%%%%%%%%%%%%%%%%%%%%%%%%%%%%%%%%%%%%%%%%%%%%%%%%%%

\section{summary}
\label{sec:sum}
In the present paper, we proposed a performance analysis scheme and a demodulation algorithm for the Kronecker channel model in MIMO wireless communication systems. For a more exact evaluation than that of our previous study using the matrix integration formula, we demonstrated that two separated manipulation steps for the product form of the channel matrix, i.e., Gaussian integration for the channel matrix and an appropriate scheme for dealing with transmitter correlation, are important for the correlated MIMO system. The numerical result for the tridiagonal correlation matrix model shows that the proposed scheme and algorithm are useful for performance analysis and for the construction of a practical demodulation algorithm.

%%%%%%%%%%%%%%%%%%%%%%%%%%%%%%%%%%%%%%%%%%%%%%%%%%%%%%%%%%%%%%%%%%%
%%%%%%%%%%%%%%%%%%%%%%%%%%%%%%%%%%%%%%%%%%%%%%%%%%%%%%%%%%%%%%%%%%%
\appendix*
\section{Perturbative expansion of free energy}

In the appendix, we derive the perturbative expressions of free energies
from two evaluations, namely, matrix integration and exact analysis.
Remember that the mutual information for a MIMO system
can be obtained by evaluating Eq. (\ref{eq:ffff1}).
For convenience, we introduce the constant $\chi$ into
the channel definition 
(see also Eqs. ({\ref{eq:FreeEnergyGeneral2}}) and (\ref{eq:submodel0})), as follows:
\begin{eqnarray}
\label{eq:model2}
\bm{r}&=&\sqrt{\chi} \bm{H}\bm{b}+ \bm{\eta},
\end{eqnarray}
where we set $\sigma=1$ in Eq. (\ref{eq:model}).
Taking QPSK modulation into account,
let us assume the probability distribution $P(b_k)$,
which satisfies the following conditions:
\begin{itemize}
\item 1. $P(b_k)$ can be factorized into the same distributions
  for the real and the imaginary parts:
   $P(b_k) = \tilde{P} ({\rm Re(b_k)}) \tilde{P}({\rm Im}(b_k))$.
\item 2. Reflection symmetry: $\tilde{P}(x)=\tilde{P}(-x)$.
\item 3. Signal power condition: 
   $\sum_{b_k} \tilde{P} ({\rm Re}(b_k)) {\rm Re}(b_k)^2
   =\sum_{b_k} \tilde{P}({\rm Im}(b_k)) {\rm Im}(b_k)^2 = 1/2$.
\end{itemize}
From these conditions, we have 
$\sum_{b_k} P(b_k) b_k^{2l-1}=0,\
 \sum_{b_k} P(b_k) |b_k|^{2l} b_k =0\ ,
 \sum_{b_k} P(b_k) |b_k|^{2}=1\ 
 (l \in \Natural) $, and so on. Quadrature-phase shift keying 
modulation is included in this case.
As mentioned in the text, we also assume that the cross-correlation
matrix $\bm H^\dag\bm H$ can be written as
$\bm H^\dag\bm H=\bm I+\rho\bm R$
with zero diagonal elements of $\bm R$, $R_{kk}=0$
and convergence of the eigenvalue distribution
of the cross-correlation matrix for $K\rightarrow \infty$.

%%%%%%%%%%%%%%%%%%%%%%%%%%%%%%%%%%%%%%%%%%%%%%%%%%%%%%%%%%%%%%%%%%%

\subsection{Expansion of free energy via matrix integration}

As shown in Section \ref{sec:analysis},
free energy is obtained via matrix integration as follows:
\begin{equation}
\overline{I_{ \bm I + \rho \bm U^{\dagger} \bm R \bm U}(\chi)}
= \extr_{\xi, \widehat{\xi}}\lb\{
 I_{\bm I}(\widehat{\xi}\chi) +
\xi\widehat{\xi} - G_{\bm I + \rho \bm R}(\xi) 
\rb\}.
\end{equation}
The function $G_{\bm I + \rho \bm R}(z)$ can be decomposed to obtain
\begin{equation}
G_{\bm I+\rho\bm R}(z)=z+G_{\bm R}(\rho z).
\end{equation}
Let us define 
$\overline{\lambda^n} \equiv \Tr(\bm R^n)/K$.
The function $G_{\bm R}(z)$ can be expressed in terms of $\overline{\lambda^n}$, as follows:
\begin{eqnarray}
G_{\bm R}(z)&=&
\frac{1}{2}\overline{\lambda^2} z^2
+ \frac{1}{3}\overline{\lambda^3}z^3
+ \frac{1}{4}\lb(\overline{\lambda^4}-2\overline{\lambda^2}{}^2
\rb) z^4
+\cdots,
\end{eqnarray}
from the formula $G(z) = \int_{0}^{z} dx ( f(x) - x^{-1})$ s.t. 
$x = \int d\lambda \rho(\lambda) ({f(x)-\lambda})^{-1}$.
Note that $\overline{\lambda}=0$ for $\bm R$ and $G(0)=0$.
Substitution into mutual information
after redefinition of $\widehat{\xi}$ yields the following:
\begin{eqnarray}
\overline{I_{ \bm I + \rho \bm U^{\dagger} \bm R \bm U}(\chi)}
&=& 
\extr_{\xi, \widehat{\xi}}\lb\{
I_{\bm I}(\chi+\widehat{\xi}\chi) +\xi\widehat{\xi} -G_{\bm{R}}(\rho \xi)
\rb\}
\nonumber \\
&=&
\extr_{\xi, \widehat{\xi}}\lb\{
I_{\bm I}(\chi)
+\widehat{\xi}\chi  I'_{\bm I}(\chi)
+\frac{(\widehat{\xi}\chi)^2}{2} I''_{\bm I}(\chi)  
+\cdots \rb.
\nonumber \\
&& \ \ \ \ \ \ \ \ \ \ \
 \lb. +\xi\widehat{\xi}
-\frac{\overline{\lambda^2}}{2}(\rho\xi)^2
-\frac{\overline{\lambda^3}}{3}(\rho\xi)^3
-\frac{\overline{\lambda^4}-2\overline{\lambda^2}^2}{4}(\rho\xi)^4+\cdots
\rb\}.
\nonumber \\
\end{eqnarray}
From the saddle point conditions with respect to $\xi,\hat{\xi}$, we have
\begin{eqnarray}
\xi&=& -\frac{\partial}{\partial \widehat\xi} I_{\bm I}
(\chi+\widehat{\xi}\chi)
= -\chi I_{\bm I}'(\chi)+ c \rho^2 + O(\rho^3),
\nonumber \\
\widehat\xi&=&
\rho G'_{\bm R}\lb(\rho\xi\rb)
=\rho G'_{\bm R}\lb(-\rho\chi I'_{\bm I}(\chi) + c \rho^3 \rb)
=\rho G'_{\bm R}\lb(-\rho \chi I'_{\bm I}(\chi)\rb)+c \rho^4
G''_{\bm R}\lb(-\rho \chi I'_{\bm I}(\chi)\rb) +O(\rho^7)
\nonumber \\
&=& -\rho^2 \overline{\lambda^2} \chi I'_{\bm I}(\chi)
+ \rho^3 \overline{\lambda^3} \chi^2 \{ I'_{\bm I}(\chi) \}^2
- \rho^4 (\overline{\lambda^4} - 2 \overline{\lambda^2}^2)
\chi^3 \{ I'_{\bm I}(\chi) \}^3 + c\rho^4 \overline{\lambda^2}
+ O(\rho^5),
\end{eqnarray} 
for $\rho$ up to the fourth order ($c$ is an $O(1)$ constant).
Substituting these equations into the original expression for the mutual information, we obtain
\begin{eqnarray}
\overline{I_{\bm I + \rho \bm U^{\dagger} \bm R \bm U}(\chi)}
&=&
I_{\bm I}(\chi)
-\frac{(\rho \chi)^2}{2} \overline{\lambda^2} \{I'_{\bm I}(\chi)\}^2
+\frac{(\rho \chi)^3}{3} \overline{\lambda^3} \{I'_{\bm I}(\chi)\}^3
-\frac{(\rho \chi)^4}{4} (\overline{\lambda^4}-2\overline{\lambda^2}{}^2)
\{I'_{\bm I}(\chi)\}^4 \nonumber \\ &&
+\frac{(\rho \chi)^4}{2}{\overline{\lambda^2}{}^2}\{I_{\bm I}''(\chi)\}
\{I_{\bm I}'(\chi)\}^2 +O(\rho^5).
\end{eqnarray}

%%%%%%%%%%%%%%%%%%%%%%%%%%%%%%%%%%%%%%%%%%%%%%%%%%%%%%%%%%%%%%%%%%%

\subsection{Expansion of rigorous free energy}

Before studying perturbative expansion of the complex 
channel, let us start with the real channel for which 
mutual communication is given by
\begin{eqnarray}
\label{eq:mutualreal}
I_{\tilde{\Rt}}^{\rm real} (\chi)
\!\! &=& \!\! -\frac{1}{K} 
 \int_{\Real^K} d\tilde{\bm r} \left\{
 \Tr_{\tilde{\bm b}} \tilde{P}(\tilde{\bm b})
   \left( \frac{\chi}{2\pi}\right)^{K/2} \!\! \exp
 \left( - \frac{\chi}{2} \left| \tilde{\bm r}-\sqrt{\tilde{\Rt}}
 \tilde{\bm b} \right|^2 \right) \right\}
 \left\{ \ln \Tr_{\tilde{\bm b}} \tilde{P}(\tilde{\bm b})
 \left( \frac{\chi}{2\pi}\right)^{K/2} \!\!
 \exp \left( - \frac{\chi}{2} 
 \left| \tilde{\bm r}- \sqrt{\tilde{\Rt}} \tilde{\bm b} \right|^2
 \right)  \right\} \nonumber \\
&& \hspace{12cm} - \frac{1}{2} \ln\left( \frac{2\pi}{\chi} \right) - \frac{1}{2},
\end{eqnarray}
where all variables and matrices are real and denoted with tilde
for discrimination between the real and the complex channels in this subsection.
By substituting $\tilde{\Rt} = \bm I + \rho \tilde{\bm R}$ into the above equation,
where $\tilde{\bm R}$ is a symmetric matrix with zero
diagonal elements, expanding with respect to $\rho$, and then performing some algebraic manipulation, we obtain the following:
\begin{eqnarray}
\label{eq:expansion}
&& I_{ \bm I + \rho \tilde{\bm R}}^{\rm real} (\chi) \nonumber \\
&=& I_{\bm I}^{\rm real} (\chi)
 + \sum_{k=1}^{4} \left. \frac{\rho^k}{k!}
 \partial_{\rho}^{k} I^{\rm real}_{\bm I + \rho \tilde{\bm R}} (\chi)
 \right|_{\rho=0}
 + O(\rho^5) \nonumber \\
&=& I_{\bm I}^{\rm real} (\chi)  
 - \frac{(\rho \chi)^2}{2!} \frac{1}{2K}
 \sum_{i_1i_2j_1j_2} \tilde{R}_{i_1 j_1}
 \tilde{R}_{i_2 j_2} 
 [\langle \tilde{b}_{i_1} \tilde{b}_{i_2} \rangle_{\rm c}
 \langle \tilde{b}_{j_1} \tilde{b}_{j_2} \rangle_{\rm c}]
 \nonumber \\
 &&
 - \frac{(\rho \chi)^3}{3!} \frac{1}{2K}
  \sum_{i_1i_2i_3j_1j_2j_3} \tilde{R}_{i_1 j_1} \tilde{R}_{i_2 j_2}
 \tilde{R}_{i_3 j_3}
 \left(
 [\langle \tilde{b}_{i_1} \tilde{b}_{i_2} \tilde{b}_{i_3} \rangle_{\rm c}
 \langle \tilde{b}_{j_1} \tilde{b}_{j_2} \tilde{b}_{j_3} \rangle_{\rm c}]
 - 2 [\langle \tilde{b}_{i_1} \tilde{b}_{j_2} \rangle_{\rm c}
  \langle \tilde{b}_{i_2} \tilde{b}_{j_3} \rangle_{\rm c}
  \langle \tilde{b}_{i_3} \tilde{b}_{j_1} \rangle_{\rm c}] 
 \right)
 \nonumber \\
 &&
 - \frac{(\rho \chi)^4}{4!} \frac{1}{2K}
  \sum_{i_1i_2i_3i_4j_1j_2j_3j_4}
 \tilde{R}_{i_1 j_1} \tilde{R}_{i_2 j_2}
 \tilde{R}_{i_3 j_3} \tilde{R}_{i_4 j_4}
 \nonumber \\
 && \hspace{-3mm} \times
 \left(
 [\langle \tilde{b}_{i_1} \tilde{b}_{i_2} \tilde{b}_{i_3}
 \tilde{b}_{i_4} \rangle_{\rm c}
 \langle \tilde{b}_{j_1} \tilde{b}_{j_2} \tilde{b}_{j_3}
 \tilde{b}_{j_4} \rangle_{\rm c}]
 -12 [\langle \tilde{b}_{i_1} \tilde{b}_{i_2} \tilde{b}_{i_3}
  \rangle_{\rm c}
  \langle \tilde{b}_{j_1} \tilde{b}_{j_2} \tilde{b}_{j_4} \rangle_{\rm c}
  \langle \tilde{b}_{i_3} \tilde{b}_{j_4} \rangle_{\rm c}]
 + 6  [\langle \tilde{b}_{i_1} \tilde{b}_{j_2} \rangle_{\rm c}
  \langle \tilde{b}_{i_2} \tilde{b}_{j_3} \rangle_{\rm c}
  \langle \tilde{b}_{i_3} \tilde{b}_{j_4} \rangle_{\rm c} 
  \langle \tilde{b}_{i_4} \tilde{b}_{j_1} \rangle_{\rm c}]
 \right) \nonumber \\
 && + O(\rho^5) \nonumber \\
 &=& I_{\bm I}^{\rm real} (\chi)  
  - \frac{(\rho \chi)^2}{4}
 \frac{\Tr \tilde{\bm R}^2}{K} [\langle \tilde{b}^2 \rangle_{\rm c}]^2
  + \frac{(\rho \chi)^3}{6}
 \frac{\Tr \tilde{\bm R}^3}{K} [\langle \tilde{b}^2 \rangle_{\rm c}]^3
 - \frac{(\rho \chi)^4}{48} 
 \frac{\sum_{i j} \tilde{R}_{ij}^4}{K} [\langle \tilde{b}^4 \rangle_{\rm c}]^2
 \nonumber \\
 && 
 - \frac{(\rho \chi)^4}{8} \left\{
 \left( \frac{\Tr \tilde{\bm R}^4}{K}
  - \frac{2 \sum_{i} ({\tilde{\bm R}^2})_{ii}^2}{K}
 + \frac{\sum_{ij} \tilde{R}_{ij}^4}{K} \right) 
  [\langle \tilde{b}^2 \rangle_{\rm c}]^4 \right.
 \left. - 2 \left( \frac{ \sum_{i} (\tilde{\bm R}^2)_{ii}^2}{K}
 - \frac{\sum_{ij} \tilde{R}_{ij}^4}{K} \right) 
  [\langle \tilde{b}^2 \rangle_{\rm c}^2] [\langle \tilde{b}^2
  \rangle_{\rm c}]^2 \right. \nonumber \\
 && \left. \hspace{15mm}
  + \frac{\sum_{ij} \tilde{R}_{ij}^4}{K}
 [\langle \tilde{b}^2 \rangle^2_{\rm c}]^2
 \right\} + O(\rho^5) \nonumber \\
 &=& I_{\bm I}^{\rm real} (\chi)  
  - \frac{(\rho \chi)^2}{4} \overline{\lambda^2}
  [\langle \tilde{b}^2 \rangle_{\rm c}]^2
  + \frac{(\rho \chi)^3}{6} \overline{\lambda^3} [\langle \tilde{b}^2
  \rangle_{\rm c}]^3
 - \frac{(\rho \chi)^4}{8} \overline{\lambda^4} [\langle \tilde{b}^2
 \rangle_{\rm c}]^4
  \nonumber \\
 && 
 - \frac{(\rho \chi)^4}{8} \left\{
  \frac{ 2  \sum_{i} ({\tilde{\bm R}^2})_{ii}^2}{K}
 (  [\langle \tilde{b}^2 \rangle_{\rm c}^2]
 - [\langle \tilde{b}^2 \rangle_{\rm c}]^2)
 [\langle \tilde{b}^2 \rangle_{\rm c}]^2
 + \frac{ \sum_{ij} \tilde{R}_{ij}^4}{K}
 \left(
  (  [\langle \tilde{b}^2 \rangle_{\rm c}^2]
 - [\langle \tilde{b}^2 \rangle_{\rm c}]^2 )^2
 + \frac{[\langle \tilde{b}^4 \rangle_{\rm c}]^2}{6}
  \right) \right\} + O(\rho^5).
 \end{eqnarray}
 Here,
 $ \langle f(\tilde{\bm b}) \rangle \equiv \Tr_{\tilde{\bm b}}
 \tilde{P}(\tilde{\bm b}) f(\tilde{\bm b})
 \exp( -\chi |\tilde{\bm r} -  \tilde{\bm b}|^2/2)
 / \Tr_{\tilde{\bm b}} \tilde{P}(\tilde{\bm b})
 \exp( -\chi |\tilde{\bm r} -  \tilde{\bm b}|^2/2),
 $ and $\langle f(\tilde{\bm b}) \rangle_{\rm c}$ is its cumulant. 
 In addition, $\langle f(\tilde{\bm b}) \rangle$ is a function of $\tilde{\bm r}$, and
 is always accompanied by the average
 $[ F(\tilde{\bm r}) ] \equiv
 (\chi / 2\pi)^{K/2}
 \int {d \tilde{\bm r}} \Tr_{\tilde{\bm b}} \tilde{P}(\tilde{\bm
 b})  F(\tilde{\bm r}) 
 \exp( -\chi |\tilde{\bm r} -  \tilde{\bm b}|^2/2) $.
where $b$ without a subscript denoted the signal at an arbitrary antenna.
 The expression in the third line of Eq. (\ref{eq:expansion}) is obtained
 by repeatedly performing integration by parts with respect to $\tilde{\bm r}$. 
 We can also show that
 $ [ \langle \tilde{b}^2 \rangle_{\rm c}] = 
  2 I^{' \rm real}_{\bm I} (\chi) $ and
 $[ \langle \tilde{b}^2 \rangle_{\rm c}^2] =
 -2 I^{''\rm real} _{\bm I} (\chi) $,
from which we have
\begin{eqnarray}
2I_{ \bm I + \rho \tilde{\bm R}}^{\rm real} (\chi) &=& 
 2I_{ \bm I}^{\rm real} (\chi)
 - \frac{(\rho \chi)^2}{2} \overline{\lambda^2}
    \{ 2 I^{' \rm real}_{\bm I} (\chi) \}^2
  + \frac{(\rho \chi)^3}{3} \overline{\lambda^3}
    \{ 2 I^{' \rm real}_{\bm I} (\chi) \}^3
 - \frac{(\rho \chi)^4}{4} \overline{\lambda^4}
    \{ 2 I^{' \rm real}_{\bm I} (\chi) \}^4
  \nonumber \\
 && \hspace{-5mm}
 - \frac{(\rho \chi)^4}{4} \left\{
 \frac{2  \sum_{i} ( \tilde{\bm R}^2 )_{ii}^2 }{K}
 (  -2 I^{'' \rm real}_{\bm I} (\chi)
     -\{ 2 I^{' \rm real}_{\bm I} (\chi) \}^2)
    \{2 I^{' \rm real}_{\bm I} (\chi)\}^2
 \right. \nonumber \\
 && \left.
 + \frac{ \sum_{ij} \tilde{R}_{ij}^4}{K}  
 \left(  
  (  -2 I^{'' \rm real}_{\bm I} (\chi)
     -\{2 I^{' \rm real}_{\bm I} (\chi)\}^2)^2
    + \frac{[\langle \tilde{b}^4 \rangle_{\rm c}]^2}{6}
  \right) \right\} + O(\rho^5).
\end{eqnarray}
Next, we convert this result into the result for a complex channel,
the mutual information of which is given by Eq. (\ref{eq:FreeEnergyGeneral}).
We can easily show that the complex channel described by
$ \bm r = \sqrt{\chi} \sqrt{\bm \Rt} \bm b + \bm z $
is equivalent to the real channel of double size
$ \tilde{\bm r} = \sqrt{\chi} \sqrt{\tilde{\bm \Rt}} \tilde{\bm b}
 + \tilde{\bm z} $, where
\begin{eqnarray}
&& 
 \tilde{\bm z} = \sqrt{2} \left(\begin{array}{c}
  \Re(\bm z) \\
  \Im(\bm z) \\
 \end{array}\right), \ \
 \tilde{\bm r} = \sqrt{2} \left(\begin{array}{c}
  \Re(\bm r) \\
  \Im(\bm r) \\
 \end{array}\right), \ \
 \tilde{\bm b} = \sqrt{2} \left(\begin{array}{c}
  \Re(\bm b) \\
  \Im(\bm b) \\
 \end{array}\right), \nonumber \\
&&  \sqrt{\tilde{\bm \Rt}} = \left(\begin{array}{cc}
  \Re(\sqrt{\bm \Rt}) & -\Im(\sqrt{\bm \Rt}) \\
  \Im(\sqrt{\bm \Rt}) & \Re(\sqrt{\bm \Rt}) \\
 \end{array}\right), \ \
  P(\bm b) = 2 \tilde{P} \left( \sqrt{2} \Re ({\bm b}) \right)
  \tilde{P} \left( \sqrt{2} \Im ({\bm b}) \right).
 \end{eqnarray}
For such a system, we can show that
$I_{\bm I} (\chi) = 2 I_{\bm I}^{\rm real} (\chi) $ and 
$I_{\bm \Rt (\chi)} =
 2 I_{\tilde{\bm \Rt}}^{\rm real} (\chi)$.
Since the eigenvalue distributions of $\bm \Rt = \bm I + \rho \bm
R$ and the corresponding 
$\bm \tilde{\Rt}$ are the same, from the relationship between the real
and complex channels, we have
\begin{eqnarray}
&& I_{\bm I + \rho \bm R}
 (\chi) = 2I_{\bm I + \rho \tilde{\bm R}}^{\rm real}
 (\chi) \nonumber \\
&=& I_{\bm I} (\chi) - \frac{(\rho \chi)^2}{2} \overline{\lambda^2}
    \{I'_{\bm I} (\chi)\}^2
  + \frac{(\rho \chi)^3}{3} \overline{\lambda^3}
    \{I'_{\bm I} (\chi)\}^3
 - \frac{(\rho \chi)^4}{4} \overline{\lambda^4}
    \{I'_{\bm I} (\chi)\}^4 \nonumber \\
&&
 - \frac{(\rho \chi)^4}{4} \left\{
  \frac{2  \sum_{i} ({\bm R^2})_{ii}^2}{K}
 \{  -I''_{\bm I} (\chi) - I'_{\bm I} (\chi)^2 \}
    \{I'_{\bm I} (\chi)\}^2
 + \frac{ \sum_{ij} (\Re (R_{ij})^4 + \Im (R_{ij})^4)}{K} 
 \left( \{-I''_{\bm I} (\chi) - I'_{\bm I} (\chi)^2 \}^2
    + \frac{C(\chi)^2}{6}
  \right) \right\} \nonumber \\
&& + O(\rho^5).
\end{eqnarray}
The function $C(\chi)$ is given by
\begin{eqnarray}
&& \hspace{1cm}
 C(\chi) \equiv
 2 (\chi / \pi)^{K}
 \int {d {\bm r}} \Tr_{\bm b} P(\bm b) 
 \langle \Re(b)^4 + \Im(b)^4 \rangle^{\rm cmp}_{\rm c} 
 \exp( -\chi | {\bm r} -  {\bm b}|^2), \nonumber \\
&& {\rm where} \ \ \langle  f(\bm b) \rangle^{\rm cmp}
 \equiv \Tr_{\bm b}
 P(\bm b) f(\bm b)
 \exp( -\chi |\bm r -  \bm b|^2)
 / \Tr_{\bm b} P(\bm b)
 \exp( -\chi |\bm r -  \bm b|^2),
\end{eqnarray}
and the subscript of the angular bracket $\rm c$ denotes the cumulant.
Substituting the definitions of $\overline{\lambda^2}$ and
$\overline{\lambda^4}$, the discrepancy between the two results is obtained as
\begin{multline}
\label{eq:discrep}
I_{\bm I + \rho \bm R}(\chi)
- \overline{I_{\bm I + \rho \bm U^{\dagger} \bm R \bm U}(\chi)}
=
-\frac{(\rho \chi)^4}{2}\frac{1}{K} \sum_{i} 
\left\{ (\bm R^2)_{ii}
- \frac{\Tr (\bm R^2)}{K} \right\}^2
\lb\{-I_{\bm I}''(\chi)- I_{\bm I}'(\chi)^2\rb\}
\{I_{\bm I}'(\chi)\}^2 
\\
-\frac{(\rho \chi)^4}{4} \frac{1}{K}
\left( \sum_{ij} \{ \Re(R_{ij})^4 + \Im(R_{ij})^4 \} \right) 
\left( \lb\{-I_{\bm I}''(\chi)- I_{\bm I}'(\chi)^2\rb\}^2
+\frac{C(\chi)^2}{6} \right)
+O(\rho^5),
\end{multline}
that is, the dominant term of the discrepancy is of the order $\rho^4$.
The factor $-I_{\bm I}''(\chi)-\{I_{\bm I}'(\chi)\}^2
=  -2 I^{'' \rm real}_{\bm I} (\chi)
     -\{2 I^{' \rm real}_{\bm I} (\chi)\}^2
=  [\langle \tilde{b}^2 \rangle_{\rm c}^2]
 - [\langle \tilde{b}^2 \rangle_{\rm c}]^2 $ 
is nonnegative, and the inequality
$\overline{I_{\bm I + \rho \bm U^{\dagger} \bm R \bm U}(\chi)}
\geq  I_{\bm I + \rho \bm R}(\chi)$ holds up to the fourth order.

%%%%%%%%%%%%%%%%%%%%%%%%%%%%%%%%%%%%%%%%%%%%%%%%%%%%%%%%%%%%%%%%%%%
%%%%%%%%%%%%%%%%%%%%%%%%%%%%%%%%%%%%%%%%%%%%%%%%%%%%%%%%%%%%%%%%%%%
\acknowledgments

The present study was supported by a Grant-in-Aid
Scientific Research on Priority Areas ``Deepening and Expansion
of Statistical Mechanical Informatics (DEX-SMI)`` from MEXT, Japan
No. 18079006. Y.K. was also supported by the JSPS Global 
COE program, ``Computationism as a Foundation for the Sciences''.

%%%%%%%%%%%%%%%%%%%%%%%%%%%%%%%%%%%%%%%%%%%%%%%%%%%%%%%%%%%%%%%%%%%


\begin{thebibliography}{99}

\bibitem{bib:Nishimori}
H. Nishimori, {\it Statistical Physics of Spin Glasses and Information
Processing: An Introduction} (Oxford University Press, Oxford, 2001).

\bibitem{bib:TanakaCD1}
T. Tanaka, Europhys Lett. {\bf 54}, 540 (2001).

\bibitem{bib:TanakaCD2}
T. Tanaka, IEEE Trans. Inf. Theory {\bf 48}, 2888 (2002).

\bibitem{bib:KabashimaDecoding}
Y. Kabashima, J. Phys. {\bf A36}, 11111 (2003).

\bibitem{bib:Guo2005}
D. Guo and S. Verd\'u, IEEE Trans. Inf. Theory {\bf 51}, 1983 (2005).

\bibitem{bib:CDMAo1}
C.-K. Wen, Y.-N. Lee, J.-T. Chen, and P. Ting,
IEEE Trans. Signal Process. {\bf 53}, 2059 (2005).

\bibitem{bib:NeirottiSaad}
J. P. Neirotti and D. Saad, Europhys. Lett. {\bf 71}, 866 (2005).

\bibitem{bib:EYSK2009}
H. Efraim, N. Yacov, O. Shental, and I. Kanter, J. Stat. Mech.:
Theory Exp. (2009) P07039.

\bibitem{bib:TUK2006}
K. Takeda, S. Uda, and Y. Kabashima, Europhys. Lett. {\bf 76}, 1193 (2006).

\bibitem{bib:THK2007}
K. Takeda, A. Hatabu, and Y. Kabashima, J. Phys. A {\bf
40}, 14085 (2007).

\bibitem{bib:mimo1} 
R. R. M\"{u}ller, D. Guo, and A. L. Moustakas, IEEE J. Select. Areas
Commun. {\bf26}, 530 (2009).

\bibitem{bib:gfunc1}
C. Itzykson and J. B. Zuber, J. Math. Phys. {\bf 21}, 411 (1980).

\bibitem{bib:gfunc2}
E. Marinari, G. Parisi, and F. Ritort, J. Phys. A {\bf 27},
 7647 (1994).

\bibitem{bib:gfunc3}
G. Parisi and M. Potters, J. Phys. A {\bf 28}, 5267 (1995).

\bibitem{bib:freeprob}
D. Voiculescu, K.J. Dykema, and A. Nica, {\it Free Random Variables}
(CRM Monograph Series, Vol.1, American Math. Society, Providence, 1992). 

\bibitem{bib:TulinoVerdu}
A. M. Tulino and S. Verd\'u, {\it Random Matrix Theory and Wireless 
Communications} (now Publishers, Hanover, 2004).

\bibitem{bib:kronecker1}
D. -S. Shiu, G. J. Foschini, M. J. Gans, and J. M. Kahn, IEEE
Trans. Commun. {\bf 48}, 502 (2000).

\bibitem{bib:diffrealcomplex}
See Appendix for the difference between the mutual informations of
the real and the complex channels (factor 2 in particular).
Note also that the functions $G$ in matrix
integration formulas for orthogonal matrix and unitary matrix
differ by overall factor 2 and factor 2 in the argument \cite{bib:gfunc2}.

\bibitem{bib:ising_Derrida} 
B. Derrida and H.J. Hilhorst, J. Phys. A {\bf16}, 2641 (1983).

\bibitem{bib:ising_Weigt} 
M. Weigt and R. Monasson, Europhys. Lett. {\bf36}, 209 (1996).

\bibitem{bib:TAP}
D. J. Thouless, P. W. Anderson, and R. G. Palmer, Philos.
Mag. {\bf 35}, 593 (1977).

\bibitem{bib:OpperWinther1}
M. Opper and O. Winther, Phys. Rev. Lett. {\bf 86}, 3695 (2001).

\bibitem{bib:OpperWinther2}
M. Opper and O. Winther, Phys. Rev. E {\bf 64}, 056131 (2001).

\end{thebibliography}
\end{document}